\documentclass[preprint]{aastex}
\pdfoutput=1
\usepackage{amsmath}
\begin{document}
\title{Synthetic Observations of Simulated AGN Jets: X-ray Cavities}
\author{P. J. Mendygral\altaffilmark{1}, S. M. O'Neill\altaffilmark{2}, \and T. W. Jones\altaffilmark{1}}
\altaffiltext{1}{Department of Astronomy, University of Minnesota, Minneapolis, MN 55455}
\altaffiltext{2}{JILA, University of Colorado, 440 UCB, Boulder, CO 80309}



\begin{abstract}
Observations of X-ray cavities formed by powerful jets from AGN in galaxy cluster cores are widely used to estimate the energy
output of the AGN.  Using methods commonly applied to observations of clusters, we conduct synthetic X-ray observations of 3D 
MHD simulated jet-ICM interactions to test the reliability of measuring X-ray cavity power.
These measurements are derived from empirical estimates of the
enthalpy content of the cavities and their implicit ages.
We explore how such physical factors as jet intermittency 
and observational conditions such as orientation of the jets with respect to
the line of sight impact the reliability of observational measurements of cavity enthalpy and age.
An estimate of the errors
in these quantities can be made by directly comparing ``observationally'' derived values with ``actual''
values from the simulations.  In our tests,
cavity enthalpy derived from observations was typically within a factor of two of the simulation values.  Cavity age and, therefore, cavity power
are sensitive to the accuracy of the estimated inclination angle of the jets. Cavity age and power estimates within a factor of two of
the actual values are possible given an accurate inclination angle.
\end{abstract}


\section{Introduction}
\label{s:intro}

X-ray images of giant cavities in galaxy clusters associated with powerful jets from central active galactic nuclei (AGN) suggest that
AGN may play an important role in the energetics of galaxy intra-cluster media (ICMs) \citep[\emph{e.g.,}][]{fabian03, birzan04, wise07}.  
The estimated minimum energy required to produce cavities is often in the range of $10^{55}$ to $10^{60}$ erg
\citep{birzan04}.  Observations of ICMs have shown the existence of a temperature floor of approximately 
2 keV \citep[\emph{e.g.,}][]{peterson02}.  The lack of gas below this temperature, 
contrary to expectations of a classical ``cooling flow'' \citep{fabian94}, 
is historically known as the ``cooling problem''.  Evidently, the energy required to suppress this cooling of the ICM below 2 keV is on the 
same order as energy in the X-ray cavities \citep{mcnamara07}.  As a result, one popular 
hypothesis that has emerged to solve the cooling problem is that energy injected into the ICM by AGN will quench cooling and 
subsequent star formation in the central cluster galaxy.  Several numerical studies such as \citet{bruggen05} and \citet{sijacki06} 
have supported this hypothesis.

X-ray cavity systems are evidently formed when low density, hot plasma originating from the AGN inflates a bubble in the ICM.  The low
density plasma produces a decrement in the line of sight intensity through the cavity from the normal ICM X-ray emission \citep{clarke97}.
The cavities produce roughly elliptical brightness depressions $\sim $20\% to 40\% below the surrounding regions \citep{mcnamara07}.  A few dozen such
cavity systems are known \citep[see][and references therein]{dong10, rafferty06}. Some cavities are filled with radio emission from relativistic particles 
and are typically found in pairs with an AGN in the cluster center between them.  Other cavities devoid of radio emission above 1.4 GHz are referred to as 
radio ghosts; \citep[\emph{e.g.,}][]{birzan08}.  
The presence of multiple cavity pairs in some cases suggests a series of outbursts from the AGN.   
Hydra A, for example, contains several attached cavities filling at least 10\% of the cluster volume
within 300 kpc of the cluster center \citep{wise07}.  Work done by \citet{morsony10}, however, suggests that the presence of multiple cavities
may be the result of the motion of a dynamic ICM.  The size of cavities varies greatly from 1 kpc in diameter for M87
\citep{young02}, for example, to over 200 kpc in diameter in Hydra A \citep{wise07}.

X-ray cavities are likely to be long lived structures, remaining intact for over 100 Myr in Hydra A, for example \citep{nulsen05}.
On the other hand, simple hydrodynamic analyses suggest cavities filled with light gas should be unstable to Rayleigh Taylor (RT) and Kelvin Helmholtz (KH) 
instabilities as they form and rise in the cluster.  Several numerical
studies have been performed, which include additional physics to stabilize the bubbles.  \citet{jones05}, for example, carried out
2D calculations of bubbles with magnetic fields finding that the fields suppress instabilities.  
\citet{reynolds05} demonstrated the stabilizing effect of a Braginskii viscosity mitigated
by Coulomb collisions.  \citet{bruggen09}
included a model for RT driven sub-grid turbulence in 3D hydrodynamic simulations of bubbles.  Their results show that turbulence can also
prevent the break up of bubbles as a by-product of resulting ICM entrainment.

Surveys of cavity systems and their energy of formation requirements have found that nearly half of the studied cavities
show evidence for sufficient power to suppress cooling for
short periods of time in their host clusters \citep{birzan04, rafferty06} if the energy
in the cavities becomes distributed in the ICM. 
There are, however, significant
uncertainties in the determination of the cavity
energy contents and associated time scales. 
The energy content, generally assumed to be measurable in terms of the
supporting pressure of the cavity, requires, for instance, accurate measurement of 
the cavity pressure and also its volume.  The cavity volume, $V$, is generally estimated from circles
or ellipses fit by eye to the cluster X-ray surface brightness distribution. That two dimensional projection is
then converted into a three dimensional ellipsoid by revolution around the long axis
\citep[\emph{e.g.,}][]{birzan04, wise07}.  
The minimum energy required to inflate a cavity containing internal energy, $E$, is taken to be the 
enthalpy content of the cavity, $E + PV$, based on the assumption that the cavity expanded 
subsonically in the ICM 
at constant pressure, $P$.
The timescale needed to inflate the cavity and to measure the associated cavity power
 is usually estimated from buoyancy or characteristic sound crossing times arguments.  

A substantial amount of effort has gone into numerical studies of outflows from AGN and 
the creation of bubbles containing hot AGN generated plasma.  To make direct comparisons with
observations, however, realistic synthetic observations from these calculations are needed.  Several authors have applied this approach
with various models of jets and bubbles in a cluster \citep[\emph{e.g.,}][]{bruggen05, diehl07, bruggen09, morsony10}.  The synthetic observations of 
magnetically dominated cavities by \citet{diehl07} were able to produce several of the observed characteristics seen in real observations including 
bright rims commonly found outlining cavities \citep{mcnamara07}.  \citet{bruggen09} were able to show consistency between their
measurement of $PV$ from synthetic observations of bubbles with sub-grid turbulence and a sample of observed cavities by \citet{diehl08}.

Studies such as \citet{dong10} and \citet{ensslin02} have tested the efficiency of detecting cavity systems from X-ray observation, but to our knowledge, 
a detailed assessment of the 
reliability of the observational techniques used to determine cavity enthalpy has not
been performed.  Synthetic observations of the complex interactions involved in the formation of X-ray cavities provide a
powerful test of these methods.
The primary goal of this paper is to test common observational techniques for 
determining cavity energetics.  We employed a pair of 3D magnetohydrodynamic (MHD) simulations of jets in realistic cluster environments
presented in \citet{oj} (henceforth, OJ10).
These simulations were post-processed to yield 
synthetic X-ray observations.  Section \ref{s:calc} describes the models and numerical methods.  Section \ref{s:synobs} and \S \ref{s:measure} 
present the observations and analysis, while \S \ref{s:conclusion} lists the conclusions of this work.  In the analysis 
we have used $H_{0}=72$ km s$^{-1}$ Mpc$^{-1}$, $\Omega_{M}=0.3$, and $\Omega_{\Lambda}=0.7$.


\section{Simulation Details}
\label{s:calc}

The simulations presented, described by OJ10, were computed on a 3D Cartesian grid using a 2nd order total variation diminishing
(TVD) non-relativistic MHD code described by \cite{rj} and \cite{ryu}.  A gamma-law
equation of state was assumed with $\gamma = 5/3$; radiative cooling was
negligible for the conditions of these simulations and was therefore ignored.
Computational details are provided in OJ10. We provide here only an outline as needed to
evaluate the present work.
The physical extent of the computational grid was $x =$ 600 kpc, $y = z =$ 480 kpc.
Each computational zone represented one cubic kiloparsec with $\Delta{x} = \Delta{y} = \Delta{z} = 1$ kpc.  Two oppositely directed jets were 
centered within the grid and aligned with the x-axis.  A passive tracer, $C_{jet}$, was advected with the flow to identify jet material from
ambient material.  Two different jet models are utilized in the present discussion; 
1) a so-called relic (RE) model in which quasi-steady jets were on for
26 Myr then turned off and, 2) an intermittent (I13) model in which the jet power
cycled on and off at 13 Myr intervals throughout the simulation.  

\subsection{Bi-directed Jet Properties}
Both simulations featured bi-directed jets that had an internal Mach 3 speed at full power,
corresponding to a physical speed $v_{jet}$ = 0.10c.  These jets
originated from a cylindrical region $r_{jet}$ = 3 kpc in radius and $l_{jet}$ = 12 kpc in length centered in the grid.  
The gas injected at
the jet origin was less dense than the ambient gas by a factor of one hundred and was initially in pressure equilibrium
with its local surroundings.  Temporal variation in the jet was controlled
by an exponential ramp in density, pressure and momentum density over 1.64 Myr for I13 and 0.65 Myr for RE.  
Physical conditions inside the jet source region were relaxed to
a volume average from a sphere surrounding the jet origin as the jet
turned off, then evolved back from instantaneous volume averages for the
local medium to the desired jet conditions as jet power resumed.
The combined power from both jets at peak was $L = 1.2\times 10^{46}$ erg s$^{-1}$. Small magnetic and 
gravitational energy contributions to the jet energy flux were ignored in
defining the jet power.  
Those energy terms were, however, followed explicitly
in the simulations and accounted for in energy exchanges between the jets and
their surroundings.

The magnetic field launched from the jet was purely toroidal, $B_{\phi} = B_{0}(r/r_{jet})$ inside
a jet core region, with $\beta = P_{jet} / (B^{2} / 8\pi) \approx 100$, on the perimeter of
the jet core. There was a thin `sheath'' surrounding the core, through which
all the jet properties, including the magnetic field transitioned to local ICM conditions.

\subsection{Cluster Environment}
\label{s:cluster}

The simulation cluster environments in OJ10 were designed to mimic a realistic,
relaxed cluster.  Gravitational potential and density
profiles were selected to yield a temperature profile typical of clusters in hydrostatic equilibrium.  A tangled ambient magnetic
field with a characteristic coherence length typical of observed clusters
was chosen to break symmetry over the grid. The local ICM magnetic
pressure averaged to about 1\% of the gas pressure, although that ratio
fluctuated by large factors over the volume.

The NFW \citep{navarro} dark matter density distribution
\begin{align}
\rho_{dm} = \frac{\rho_{s}}{\left(\frac{r}{r_{dm}}\right)\left(1 + \frac{r}{r_{dm}}\right)^{2}},
\end{align}
was used to generate the gravitational acceleration
\begin{align}
g(r) = -\frac{4\pi Gr_{dm}^{3}\rho_{s}}{r^{2}}\left[ln\left(1 + \frac{r}{r_{dm}}\right) - \frac{r}{r + r_{dm}}\right] \label{f:grav}
\end{align}
where $r_{dm} =$ 400 kpc and $\rho_{s} \approx 4.3 \times 10^{-26}$ g cm$^{\textrm{-3}}$.  This gave a virial mass of $M_{v} = 5 \times 
10^{14}\,\,M_{\odot}$ for a virial radius of 2 Mpc, which was within a factor of a few of the Perseus Cluster \citep[\emph{e.g.,}][]{ettori98}.

The gas density of the cluster was initialized with a density distribution
\begin{align}
\rho_{a}(r) = \rho_{0} \left[ \frac{f_{1}}{\left(1 + \left(\frac{r}{r_{c1}}\right)^{2}\right)^{\frac{3\beta}{2}}} + \frac{f_{2}}{\left(1 + 
\left(\frac{r}{r_{c2}}\right)^{2}\right)^{\frac{3\beta}{2}}}\right], \label{rhobeta}
\end{align}
given by OJ10, where $f_{1} = 0.9$, $f_{2} = 0.1$, $r_{c1} =$ 50 kpc, $r_{c2} =$ 200 kpc and $\beta =$ 0.7.  The density scale was 
$\rho_{0} = 8.33 \times 10^{-26}$ g cm$^{-3}$.  The pressure
was determined by hydrostatic equilibrium, yielding a temperature profile resembling typical clusters (cf OJ10).  
The central pressure was  $P_{0} = 4 \times 10^{-10}$ dyne cm$^{-2}$ giving a sound speed $c_{0} = 895$ cm s$^{-1}$ in the cluster core
for a $\gamma =$ 5/3 gas.

On top of the initialized hydrostatic equilibrium in the ICM, a Kolmogorov spectrum of density fluctuations was imposed 
with a maximum local amplitude, $\pm0.10\rho_{a}(r)$, as described by OJ10.

The initially tangled and divergence-free cluster magnetic field was given by OJ10 as
\begin{align}
\overrightarrow{B} = B_{\theta}\hat{\theta} + B_{\phi}\hat{\phi}
\end{align}
where the components are
\begin{align}
B_{\theta} = \frac{F_{1}(r)\cdot m}{r}sin\theta\,cos(m\phi) \\
B_{\phi} = \frac{F_{2}(2)\cdot n}{r}sin(n\theta) - \frac{F_{1}(r)}{r}sin(m\phi)\,sin(2\theta)
\end{align}
with $m = n =$ 3.  $F_{1}(r)$ and $F_{2}(r)$ are functions designed to keep an approximately constant $\beta$ atmosphere with fluctuations that
vary over scales of a few tens of kpc.  The scale of the fields maintains a $\beta \approx$ 100 on average over the cluster volume.  The 
maximum magnitude of the field is $\sim 10 \mu$G.

\subsection{Relativistic ``Cosmic Ray'' Electrons}
\label{s:crs}

The simulations included a population of relativistic Cosmic ray electrons (CRs) passively advected with the MHD quantities.  
The numerical details of the CR transport are given in 
\citet{jones99}, \citet{treg01} and \citet{treg04}.  A small, fixed fraction of
the thermal electron flux through shocks was injected into the CR electrons population and 
subjected to first order Fermi acceleration according to the standard test-particle
theory. Downstream of shocks the CRs were also subject to adiabatic and synchrotron/inverse Compton radiative
energy changes.
The nominal CR pressure, which was neglected, was generally less than 1\% of the gas pressure.
CRs with Lorentz factors from $\gamma =$ 10 to $\gamma \sim 1.6\times10^{5}$ were tracked as a 
piecewise power law distribution.

The inclusion of CRs allowed us to calculate inverse Compton and synchrotron emissions in a self-consistent manner.  A separate analysis paper
will include detailed consideration of radio synchrotron emission and high energy non-thermal
X-ray ($>$ 10 keV) emission.
Only X-rays below 10 keV are discussed in this paper. Those are entirely dominated in our
computations by thermal emissions from the cluster ICM, although inverse
Compton emissions are included.


\section{Synthetic X-ray Observations}
\label{s:synobs}

The physical quantities evolved through MHD simulations of radio jets provide unparalleled intuition into the complex dynamics of MHD flows,
but it has not always been intuitive to relate these quantities to observation.  To make this connection and address questions raised
from observations, emission processes in these simulations must be properly calculated and converted into a synthetic observation.

The approach used here to model synthetic X-ray observations was based on \citet{treg02}.  
Observations were computed for an assumed cluster redshift, $z = 0.0594$, ($D_L = 240$ Mpc)
corresponding approximately to the Hydra cluster \citep{wise07}.
In order to understand better the influence of projection effects, we
carried out the synthetic observations at three representative angles, 
$i = 80^{\degr}, 45^{\degr}, ~{\rm and}~30^{\degr}$, between
the jet axis and the line of sight.
Two emission mechanisms were included; thermal bremsstrahlung and inverse Compton scattering off of CMB photons.  
In each zone of the
computational grid we calculated emissivities based on local properties of the thermal or CR
electron population, then corrected them for the cluster redshift.  

Thermal bremsstrahlung or free-free emissivity was computed as  
\begin{align}
j_{\nu_{local}} = 5.4\times10^{-39}\,g_{ff}(\nu_{local},T_{e})Z_{i}^{2}\frac{n_{e}n_{i}}{T_{e}^{1/2}}e^{-h\nu_{local}/kT_{e}}
\,\,erg\,cm^{-3}\,s^{-1}\,sr^{-1}, \label{f:brem}
\end{align}
where  $\nu_{local} = \nu_{obs}\left(1 + z\right)$.
The free-free Gaunt factor, $g_{ff}$, was computed by interpolation from the values calculated for plasma with typical 
ICM properties in Table 1 of \citet{nozawa}.
We assume a fully ionized $Z_{i} = 1$ hydrogen gas with an ideal gas equation of state
where the average temperature per zone was 
$T_{e} = T_{i} = T(keV)= \mu P m_{H}/(1.602\times10^{-9}\rho )$  
with $\mu =$ 1/2 and $P$ and $\rho$ in cgs units. The numerical resolution of 
discontinuities in the simulations was a few zones.  
Consequently, the contact discontinuity between AGN (jet) and ICM plasmas was a few zones of moderately 
high density, very high temperature gas.  These transition regions were artificial and should not in 
the absence of some equivalent, real viscous mixing, contribute to line of sight intensities.  To reduce this artifact,
any zone with $C_{jet} \geq 0.01$ (partially AGN plasma) had $j_{\nu_{local}}$ for
thermal bremsstrahlung set to zero.  Equation \ref{f:brem} was integrated numerically over a given range of frequencies to simulate finite bandwidths 
of real instruments.  
This paper focuses on energy ranges accessible to observatories typified by \emph{Chandra}. At those energies the inverse Compton emission in these 
simulations is negligible.  
Consequently, we omit details of their computation.

Assessing projection effects is critical to comparisons of these synthetic observations with real observations.  We developed a parallelized
ray casting engine that allows the user to define an arbitrary orientation and resolution for the output images.  A ray was cast normal to 
the image plane through the appropriately aligned grid of emissivities.  Tri-linear interpolation was used at regular intervals 
along the ray and summed to give the total intensity along the line of sight, assuming an optically thin medium.  Finally, intensities were 
converted into fluxes per pixel by multiplying the line-of-sight intensity by the solid angle of an image pixel.  
The image resolution was set to 1 arc sec, which matched the
simulation 1 kpc physical resolution at the selected 240 Mpc source distance.

\subsection{Relic (RE) Observations}
\label{s:REobs}

\begin{figure}
  \includegraphics[height=.8\textheight]{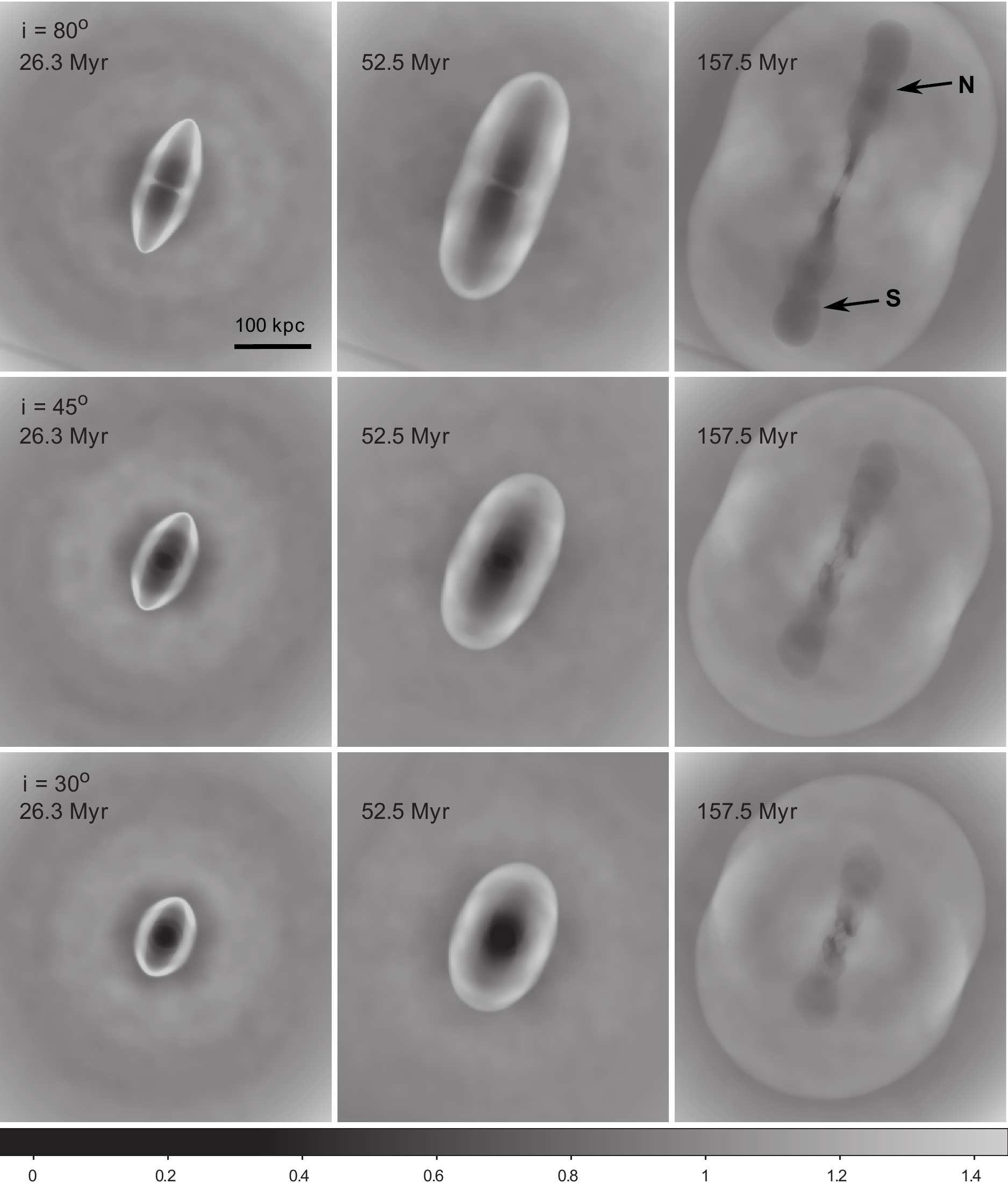}
  \caption{Synthetic observations of the RE simulation at three epochs and three orientations from the energy band 1.5-2.5 keV.  
Each row corresponds to one inclination
angle with the time progressing from left to right.  The observations were divided by a best-fit double $\beta-$profile to emphasize the X-ray cavities.  
The \textbf{N} and \textbf{S} cavity labels distinguish the cavities from each jet.}
\label{fig:REobs}
\end{figure}

Synthetic X-ray observations of the RE, relic simulation are shown for several
times and projection orientations in Figure \ref{fig:REobs}.
Following a common practice designed to highlight AGN-blown cavities, the computed
brightness distribution of the ICM outside of the identified cavities was fit with a double 
$\beta$-profile (\S \ref{s:enthalpy})
and then divided out to accentuate the X-ray cavities.
The double $\beta$-profile was determined independently for each synthetic X-ray observation.
Figure \ref{fig:REobs} shows the synthetic X-ray observations
in a 1.5-2.5 keV band divided by the best-fit 
double $\beta-$profile in each instance.  Time evolution of the system is displayed
from left to right.  At the
earliest time shown, 26.3 Myr, the jets had just turned off.
Each row corresponds to a different orientation, with the inclination angle of the
jet with respect to the line of sight decreasing from top to bottom.  
There are several
notable features in each observation.  
Cavities are seen as brightness decrements from the surrounding emission.  A pair of cavities
is seen at 26.3 and 52.5 Myr at large inclination but appear to merge into a single cavity at small inclination.  Presumably X-ray emission from a 
central galaxy would prevent the two cavities from appearing as a single cavity.  Our simulations, however, do not include 
emission from gas bound specifically to the central galaxy.  All inclinations show a pair of
cavities at 157.5 Myr.  The contrast in brightness of the cavity to the surrounding gas diminishes with both distance from cluster center and
decreasing inclination.  A detailed discussion on these trends can be found in \citet{ensslin02}.  The bow shock from the jets is 
seen in all observations.  At early times the bow shock appears as a bright rim surrounding the cavities.  At later times the bow shock 
has moved far from the edge of the cavities and no longer appears as a bright rim.

\subsection{Intermittent (I13) Observations}
\label{s:I13obs}

Figure \ref{fig:I13obs} shows a similar set of observations to those in Figure \ref{fig:REobs},
but for the I13, intermittent jet simulation.  Several new features are seen
with the introduction of jet intermittency.  Late times at every inclination reveal
``ripples'' between the cavities and the bow shock.
These features correspond to sound waves generated at the cavity walls during periods of jet activity.  
Similar ``ripples'' have been seen in
observations of the Perseus cluster (\cite{fabian03}).  A second, related
distinction from the RE 
observations is the appearance of bright rims outlining the cavities at every
epoch.  At smaller inclination angles the bright rims resemble the ``arms'' seen on smaller scales in NGC 4636 \citep{baldi}.


\section{Cavity Measurements}
\label{s:measure}

In the following analysis we attempted to apply common techniques for extracting two fundamental parameters for each cavity detected in
each observation throughout the elapsed time of each jet model.  Every epoch for a given simulation represents a separate test for measuring both
cavity enthalpy and cavity age.  Following this time evolution allows us to detect biases and trends in the quality of the measurements.  For the
remainder of this paper we refer to values measured directly from the simulation
data as the ``actual'' values, while values measured from the synthetic
observations are referred to as ``observed'' values.  We report the fractional error on a measured quantity $x$ 
as $\epsilon_{x} \equiv (x_{observed} - x_{actual})/x_{actual}$ for the remainder of this paper.

\subsection{Enthalpy}
\label{s:enthalpy}

The minimum energy required to produce a cavity is generally
estimated as the total thermal energy in the cavity and the work done inflating the cavity slowly at constant pressure; that is, 
the enthalpy in the cavity, $H = U_{therm} + PV\sim {\rm several}\times PV$. 
In particular, if the adiabatic index of the cavity plasma is $\gamma_c$,
\begin{align}
H = \frac{\gamma_c}{\gamma_c - 1}PV.
\end{align}
For a gas with $\gamma_c = 5/3$, applicable to our simulations, this gives $H = (5/2)PV$.  
Estimation of cavity enthalpy, under the assumption that the cavity was inflated at 
its current location, requires knowledge of both the cavity volume and surrounding gas pressure. 
Since the AGN activity disturbed large volumes of the ICM it is not straightforward to
determine either its pressure distribution or, for that matter, the volume occupied by
the AGN generated cavity. A common strategy to resolve these two problems involves
fitting a simple, symmetric brightness profile to regions of emission that seem
not to include cavity structures. That profile can then be used to
obtain estimates for the average radial ICM properties. 
There are several variations of this strategy \citep[\emph{e.g.,}][]{wise04, birzan04}. 
Our goal was not to determine the best strategy but to use a common approach as an example.  
We followed a procedure similar to \citet{wise04} and \citet{xue00} to extract pressure and \cite{birzan04} to extract volume from each observation.  

\begin{figure}
  \includegraphics[height=.28\textheight]{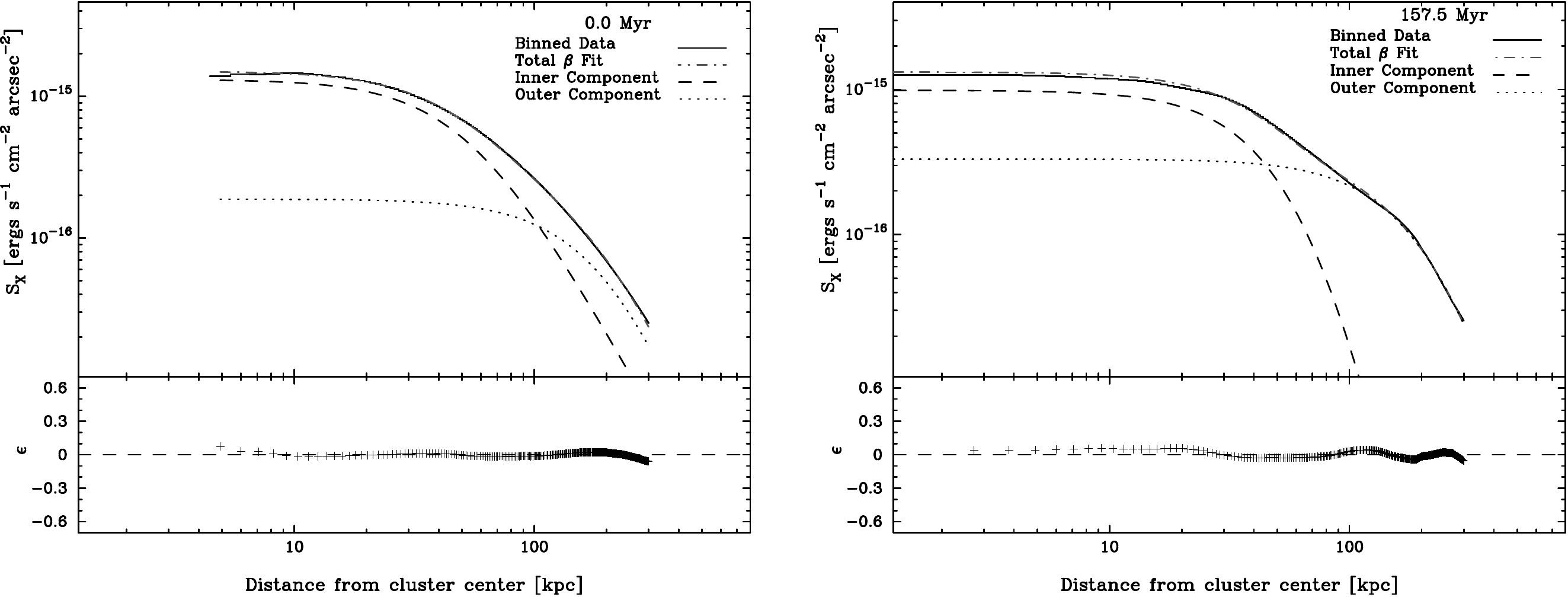}
  \caption{Double $\beta-$profile fits to azimuthally averaged radial brightness brightness distributions measured from the synthetic X-ray observations.  
On the \emph{left} is the fit for the initialized cluster of both RE and I13 observed at an inclination of 80$^{\text{o}}$.  On the \emph{right} is a 
fit for the RE simulation at 157 Myr observed at an inclination of 45$^{\text{o}}$.  The bottom panel of each plot shows the fractional error of the fit 
to the observed distribution.}
\label{fig:betafit}
\end{figure}

For this exercise we used a double $\beta-$profile profile \citep[\emph{e.g.,}][]{ikebe96} of the form
\begin{align}
S_{X}(r_{p}) = S_{0} \left( S_{01}\left[1 + \left(\frac{r_{p}}{R_{C1}}\right)^{2}\right]^{1/2 - 3\beta_{1}} +
           S_{02}\left[1 + \left(\frac{r_{p}}{R_{C2}}\right)^{2}\right]^{1/2 - 3\beta_{2}} \right), \label{f:dbeta}
\end{align}
where $r_{p}$ is the projected distance from cluster center, to model the brightness distribution of the X-ray emitting ICM.  \citet{xue00} 
discuss the benefits of using this profile as opposed to a single 
$\beta-$profile.  The profile was fit independently to each 1.5-2.5 keV  
synthetic X-ray observation of the RE and I13 simulations.
The synthetic images were divided into annular bins, each $\approx$ 1 arc sec in width.  To remove any effects of the X-ray cavities in
characterizing the brightness profile of the cluster plasma, a set of ellipses was chosen that best fit each cavity by eye.  Any pixels within 
these ellipses 
were excluded from the annular bins.  The average flux from the remaining pixels was used to define an azimuthally averaged brightness profile that was 
fit with the double $\beta$-profile.  Refer to Appendix \ref{a:fitting} for details regarding the fitting procedure.  Figure \ref{fig:betafit} 
shows example double $\beta-$profile fits for observations of both models
at an inclination of $i$ = 45$^{\text{o}}$.  The best fit profiles resulted in
$0.5 \le \beta_{1} \le 1.5$, $0.9 \le \beta_{2} \le 1.8$ with typical values $R_{C1} \sim 50$ kpc, $R_{C2} \sim 200$ kpc, $S_{01} \sim 0.8$, and
$S_{02} \sim 0.2$.  The undisturbed cluster parameters were $\beta_{1} = 0.7$, $\beta_{2} = 1$, $R_{C1} = 55$ kpc, $R_{C2} = 260$ kpc, $S_{01} = 0.9$,
and $S_{02} = 0.1$.

\subsubsection{Cluster Temperature Profile}
\label{s:temp}

\begin{figure}
  \includegraphics[height=.4\textheight]{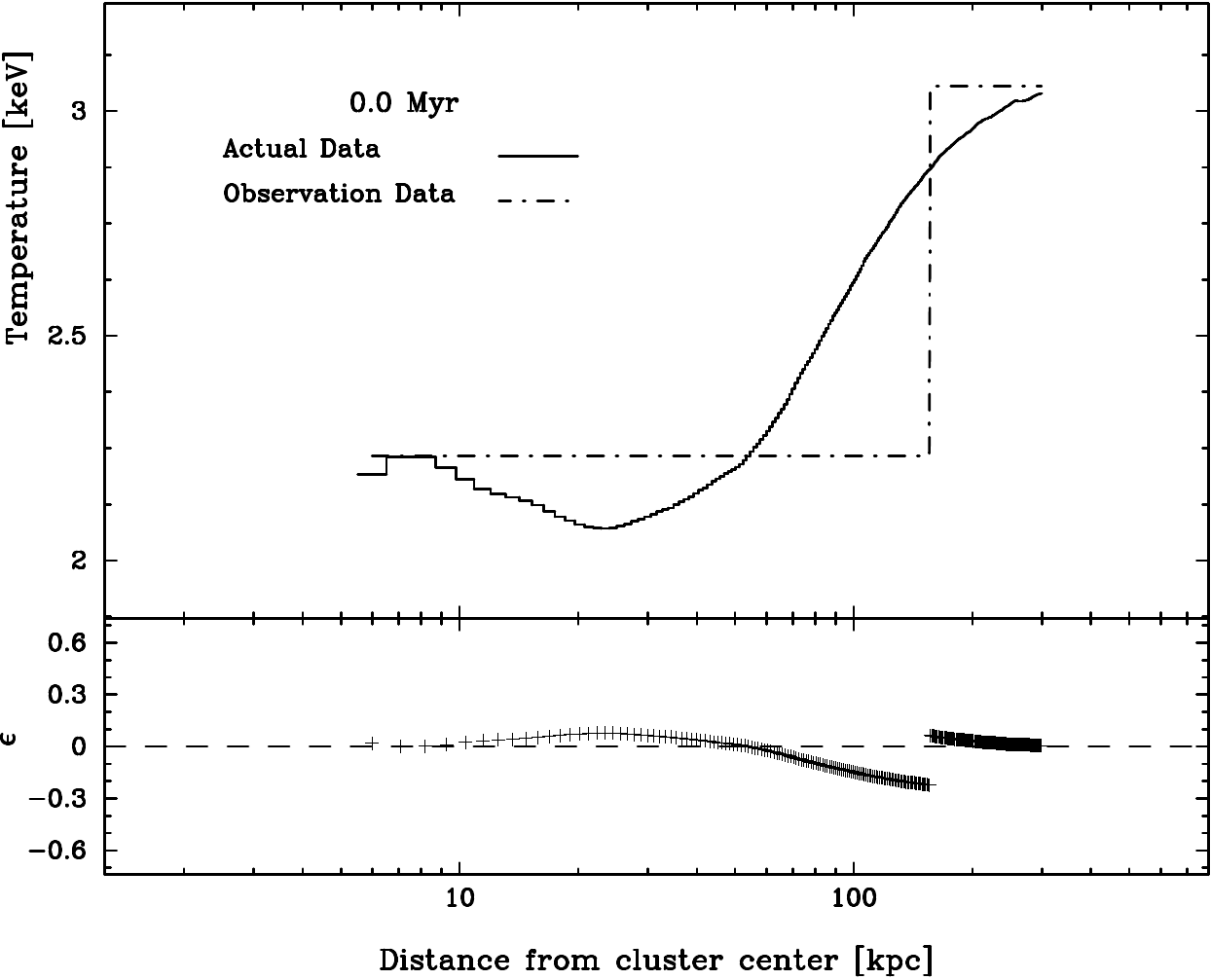}
  \caption{The observed and actual temperature profiles from the initialized cluster of both the RE and I13 simulations.
The bottom shows the fractional error of the observationally derived profile from the actual profile.}
\label{fig:temp}
\end{figure}

The ICM temperature, $T_{ICM}$, at a given projected radius, $r_{p}$, was determined from the ratio of fluxes in two bands; 
1.5-2.5 keV and 9.5-10.5 keV. 
In particular, the equation
\begin{align}
\frac{S_{X,1.5-2.5}(r_{p})}{S_{X,9.5-10.5}(r_{p})} = \frac{\int_{\nu = (1+z)1.5\,keV/h}^{\nu = (1+z)2.5\,keV/h}g_{ff}(\nu,T_{ICM})e^{-h\nu/T_{ICM}}d\nu}
{\int_{\nu = (1+z)9.5\,keV/h}^{\nu = (1+z)10.5\,keV/h}g_{ff}(\nu,T_{ICM})e^{-h\nu/T_{ICM}}d\nu} \label{f:temperature}
\end{align}
was solved for $T_{ICM}$ using the aforementioned double $\beta$-profile fits.
Following \citet{wise04}, we assumed that the two components 
of the double $\beta$-profile corresponded to two phases of the ICM with temperatures $T_{ICM,1}$ for the inner component and $T_{ICM,2}$ for the
outer component.  $T_{ICM,1}$ was taken to be the minimum and $T_{ICM,2}$ the maximum temperatures found using Equation \ref{f:temperature}.  The 
projected radius for the transition from $T_{ICM,1}$ to $T_{ICM,2}$ was chosen to be the average of $R_{C1}$ and $R_{C2}$.  Figure 
\ref{fig:temp} shows a comparison between the actual azimuthally averaged temperature profile as a function of physical radius from the RE initial 
conditions and the two component 
projected profile.  Note that $T_{ICM,1}$ mostly exceeded the actual inner core temperatures, since hotter gas along the line of sight contaminated 
$S_{X}$ at small $r_{p}$.

\subsubsection{Cluster Electron Density Profile}
\label{s:density}

The radial thermal electron density profile of component $i=1,2$ was obtained by inverting equation \eqref{f:dbeta}, following the derivation of 
\citet{xue00}; 
\begin{align}
n_{ei}^{2}(r_{p}=0) = \left(\frac{4\pi^{1/2}}{\alpha(T_{ICM,i})g_{i}\mu_{e}}\right)\left(\frac{\Gamma(3\beta_{i})}{\Gamma(3\beta_{i} - 1/2)}\right)
\left(\frac{S_{0i}}{R_{Ci}}\right)A_{ij},
\end{align}
where
\begin{align}
\alpha(T_{ICM,i}) = \frac{2^{4} e^{6}}{3 m_{e} \hbar c^{2}} \left(\frac{2\pi\,1.602\times10^{-9}\,T_{ICM,i}}{3 m_{e} c^{2}}\right)^{1/2}, \\
g_{i} = \int_{\nu = (1+z)1.5\,keV/h}^{\nu = (1+z)2.5\,keV/h} g_{ff}(\nu,T_{ICM,i})\,e^{-h\nu/T_{ICM}}\,d\nu,
\end{align}
and
\begin{align}
\frac{1}{A_{ij}} = 1 + \frac{R_{Ci}S_{0j}g_{i}}{R_{Cj}S_{0i}g_{j}}\left(\frac{T_{ICM,i}}{T_{ICM,j}}\right)^{1/2}
                   \left[\frac{\Gamma(3\beta_{j})\Gamma(3\beta_{i} - 1/2)}{\Gamma(3\beta_{i})\Gamma(3\beta_{j} - 1/2)}\right], \nonumber \\
j = 1,2\,\,and\,\,j\ne i.
\end{align}
The values for $S_{0i}$, $R_{Ci}$, and $\beta_{i}$ were the best fit values for each component from the 1.5-2.5 keV observation.  
For simplicity, we assumed pure hydrogen.  The electron weight, $\mu_{e} = 2 / (1 + X)$, where $X$, the hydrogen mass fraction, was therefore unity.  
The total electron density at a projected radius $r_{p}$ was determined by
\begin{align}
n_{e}(r_{p}) = \displaystyle\sum_{i=1}^{2}n_{ei}(r_{p}) = \left(n_{e}(r_{p}=0) \displaystyle\sum_{i=1}^{2}n_{ei}(r_{p}=0)\left[1 + 
       \left(\frac{r_{p}}{R_{Ci}}\right)^{2}\right]^{-3\beta_{i}}\right)^{1/2}. \label{f:betaden}
\end{align}

\begin{figure}
  \includegraphics[height=.55\textheight]{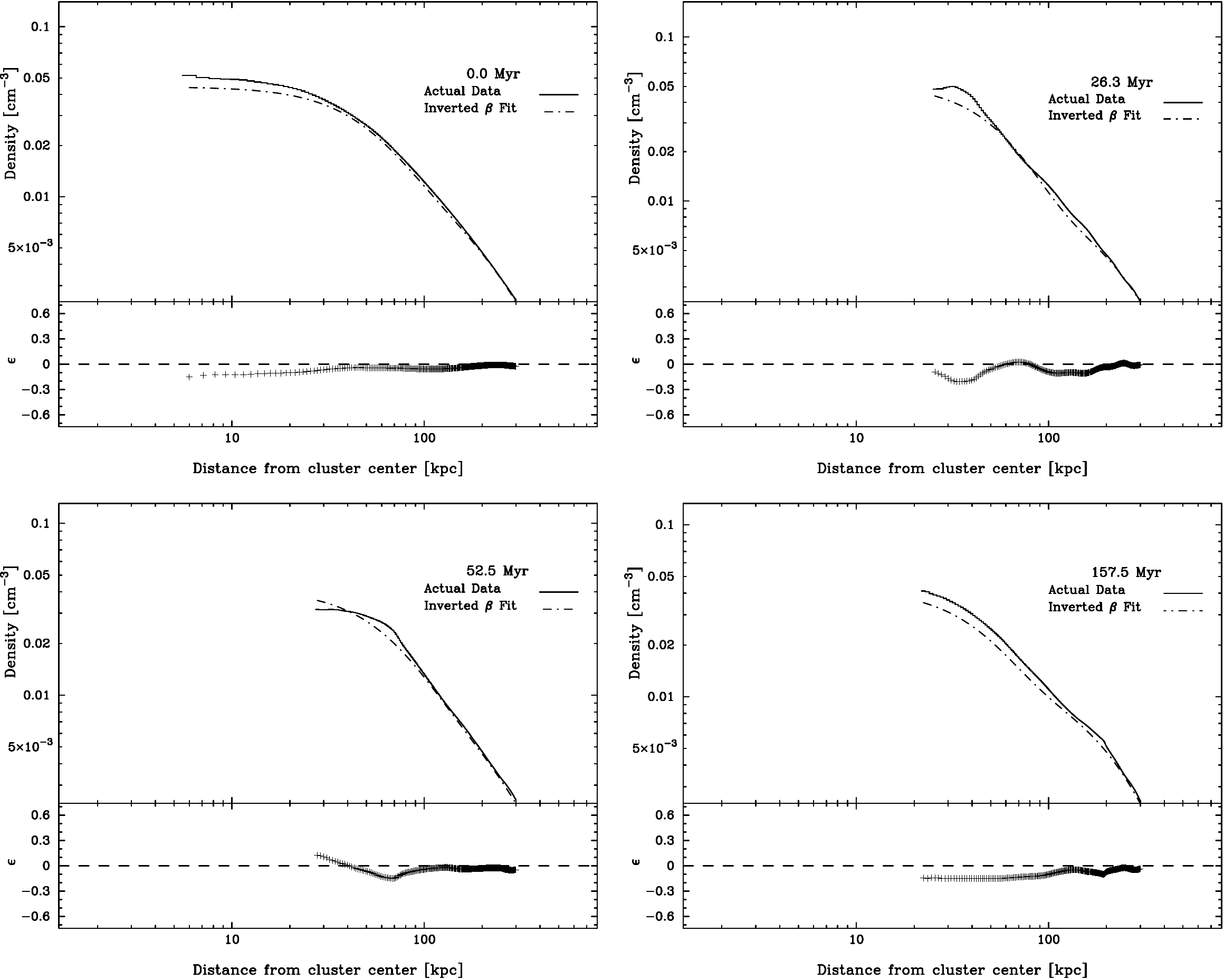}
  \caption{ICM electron density profile for the RE simulation at four characteristic epochs at an inclination of 45$^{\text{o}}$.  The \emph{solid line} 
shows the actual azimuthal average measured
from the simulation.  The \emph{dash-dot line} is the profile determined from the inverted double $\beta$-profile fits to the synthetic X-ray observations.
The bottom of each panel shows the fractional error of the observed profile from the actual profile.}
\label{fig:REdensity}
\end{figure}

\begin{figure}
  \includegraphics[height=.55\textheight]{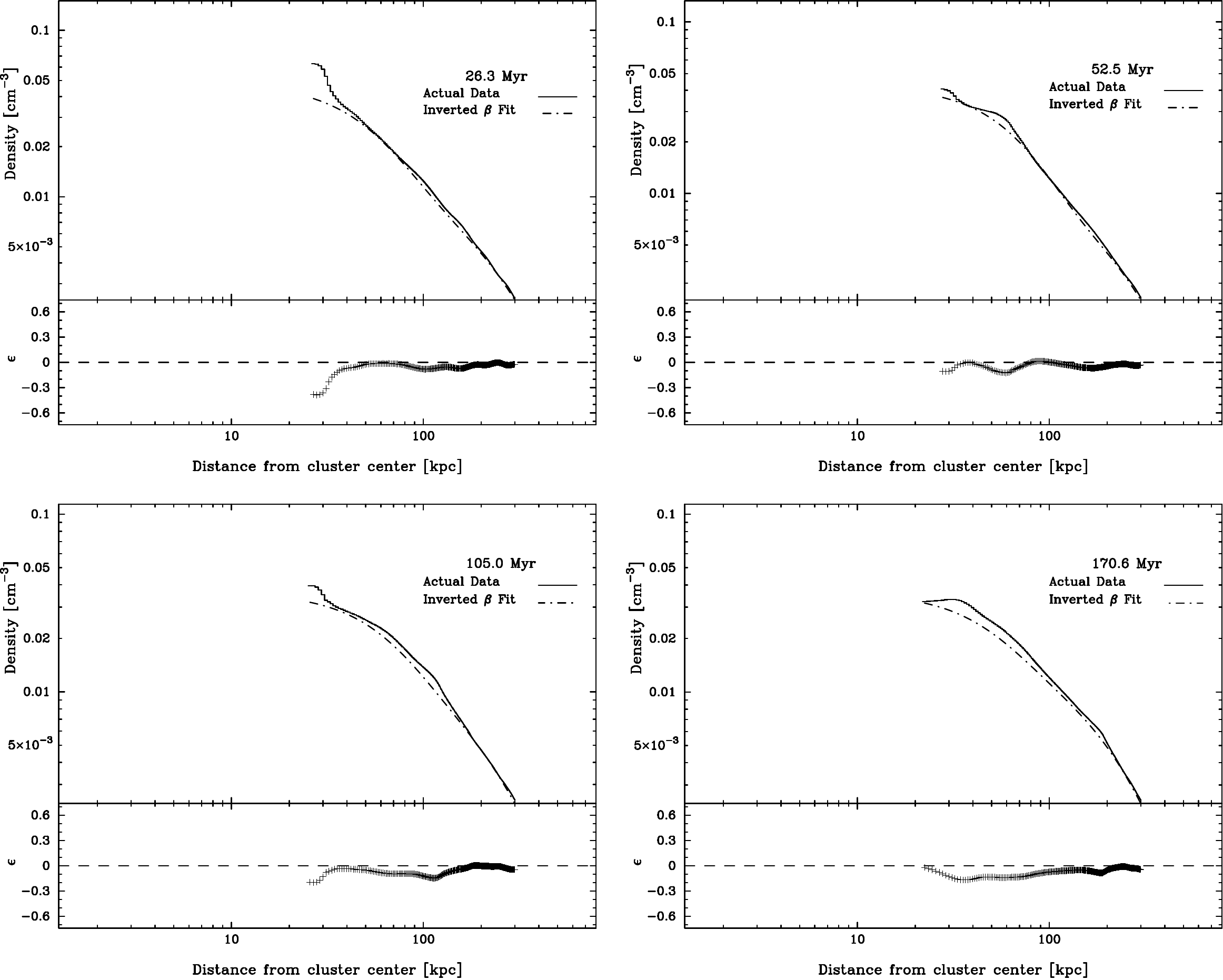}
  \caption{Similar to Figure \ref{fig:REdensity} for the I13 simulation.}
\label{fig:I13density}
\end{figure}

Figures \ref{fig:REdensity} and \ref{fig:I13density} show example observed electron density profiles determined by Equation \ref{f:betaden} compared 
to the actual azimuthally averaged electron density from the RE simulation from observations at $i$ = 45$^{\degr}$.  The data for the actual density 
were generated considering only computational zones with $C_{jet}$ = 0 
(pure ICM plasma) to avoid any contamination by AGN plasma.  Near to and within the jet launching region, $\lesssim\,l_{jet}$, this 
condition was, of course, not met while the jets were active.  For this reason there are no data points at small radii in Figures \ref{fig:REdensity} and 
\ref{fig:I13density} when the jets were active.

The initial conditions in the upper left panel of Figure \ref{fig:REdensity} reveal a bias towards 
lower density within the inner core radius (${\lesssim}$ 50 kpc) with a fractional error $\epsilon_{\rho} \sim$ 15\%.  
This is a result of the bias toward
higher temperatures in the determination of $T_{ICM,1}$ discussed in \S \ref{s:temp}.  By holding $j_{\nu}$ constant in Equation \ref{f:brem} for 
an observation with 
$h\nu_{local}$ approximately equal to the actual gas temperature it can be shown that an overestimate of the gas temperature will result in an
underestimate of the electron density.  After jet activity terminated in the RE simulation, the observed density profile matches the actual ICM 
profile with  an error, $\epsilon_{\rho} \sim$ 30\%.  The largest contribution to the
error is from regions influenced by the bow shock.  Excluding these regions gives $\epsilon_{\rho} \lesssim$ 15\%.  Evidence of the bow shock is seen 
in the actual density profile as a bump in excess of the smooth,
observationally determined profile at $r_{p}$ = 35 kpc for 26.3 Myr, 70 kpc for 52.5 Myr, and 200 kpc at 157.5 Myr.  These distances correspond to the
cluster-centric distance of the bow shock orthogonal to the jet axis.
In general, the observed distribution closely matches the actual distribution for radii outside the inner core radius, $R_{C1}$, at all times also with
$\epsilon_{\rho} \lesssim$ 10\%.  

The I13 profiles in Figure \ref{fig:I13density}, similarly display fractional errors for the observed distribution
 $\epsilon_{\rho} \lesssim$ 15\% at all times exterior to $R_{C1}$.  Evidence of the bow shock is seen here as well in the actual profile as a bump 
in excess over the observed profile at $b$ = 20 kpc for 26.3 Myr, 55 kpc for 52.5 Myr, 100 kpc for 105 Myr, and 200 kpc for 170.6 Myr.
The electron density profile of the ICM obtained from the brightness profile was reliable to within $\sim$ 20\% outside of regions influenced
by shocks regardless of jet intermittency and observed inclination.  Inside of shock influenced regions the fractional error was as high as $\sim$ 40\%.

\subsubsection{Cluster Pressure Profile}
\label{s:pressure}

\begin{figure}
  \includegraphics[height=.55\textheight]{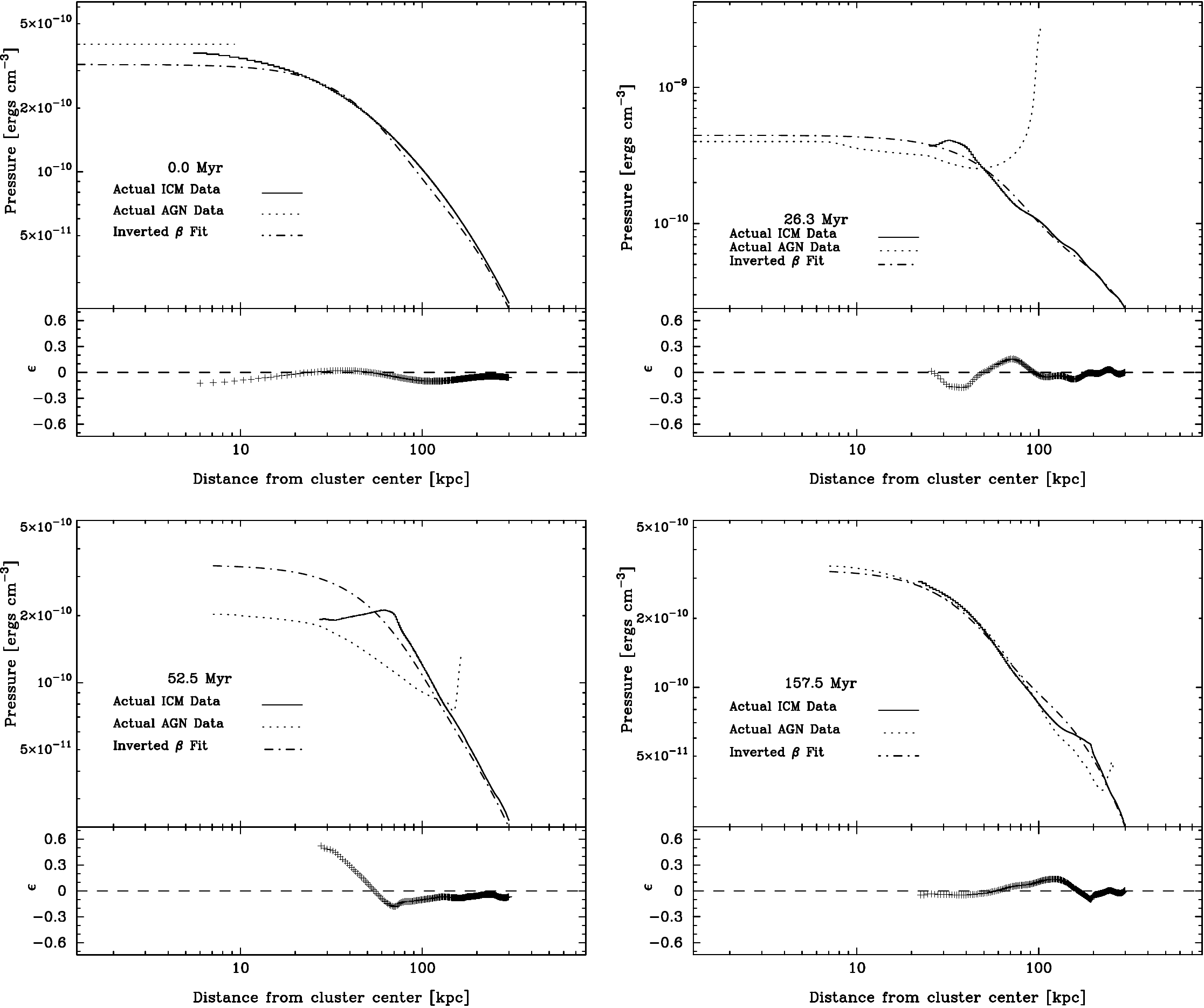}
  \caption{ICM pressure profile for the RE simulation at four characteristic epochs at an inclination of 45$^{\text{o}}$.  The \emph{solid line} shows the 
actual azimuthal average measured
from the simulation.  The \emph{dash-dot line} is the profile determined from the inverted double $\beta$-profile fits to the synthetic X-ray observations.
The dotted line is the average pressure in AGN plasma at that radius measured from the simulation.
The bottom of each panel shows the fractional error of the observed profile to the actual ICM profile.}
\label{fig:REpressure}
\end{figure}

\begin{figure}
  \includegraphics[height=.55\textheight]{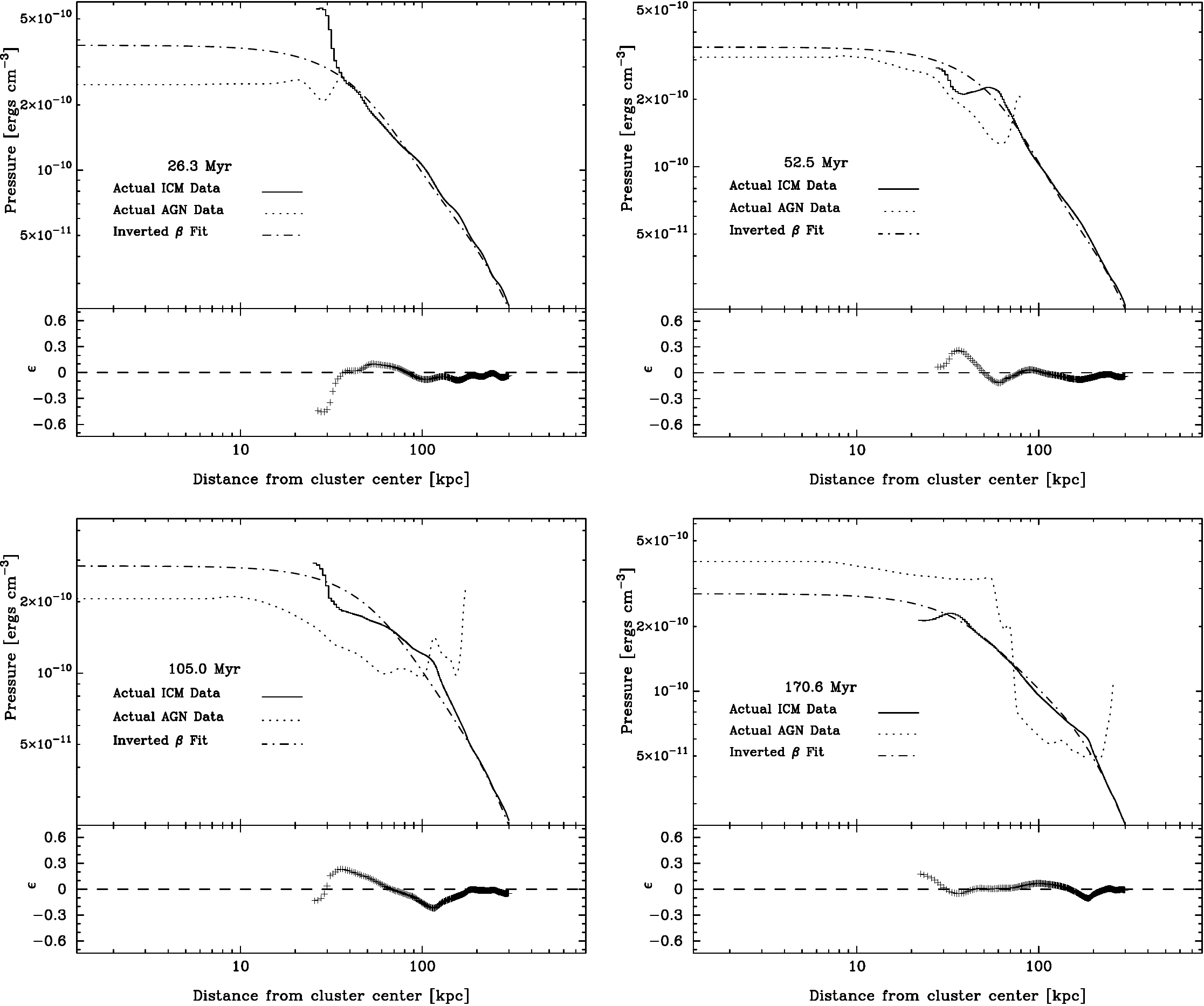}
  \caption{Similar to Figure \ref{fig:REpressure} for the I13 simulation.}
\label{fig:I13pressure}
\end{figure}

\begin{figure}
  \includegraphics[height=.6\textheight]{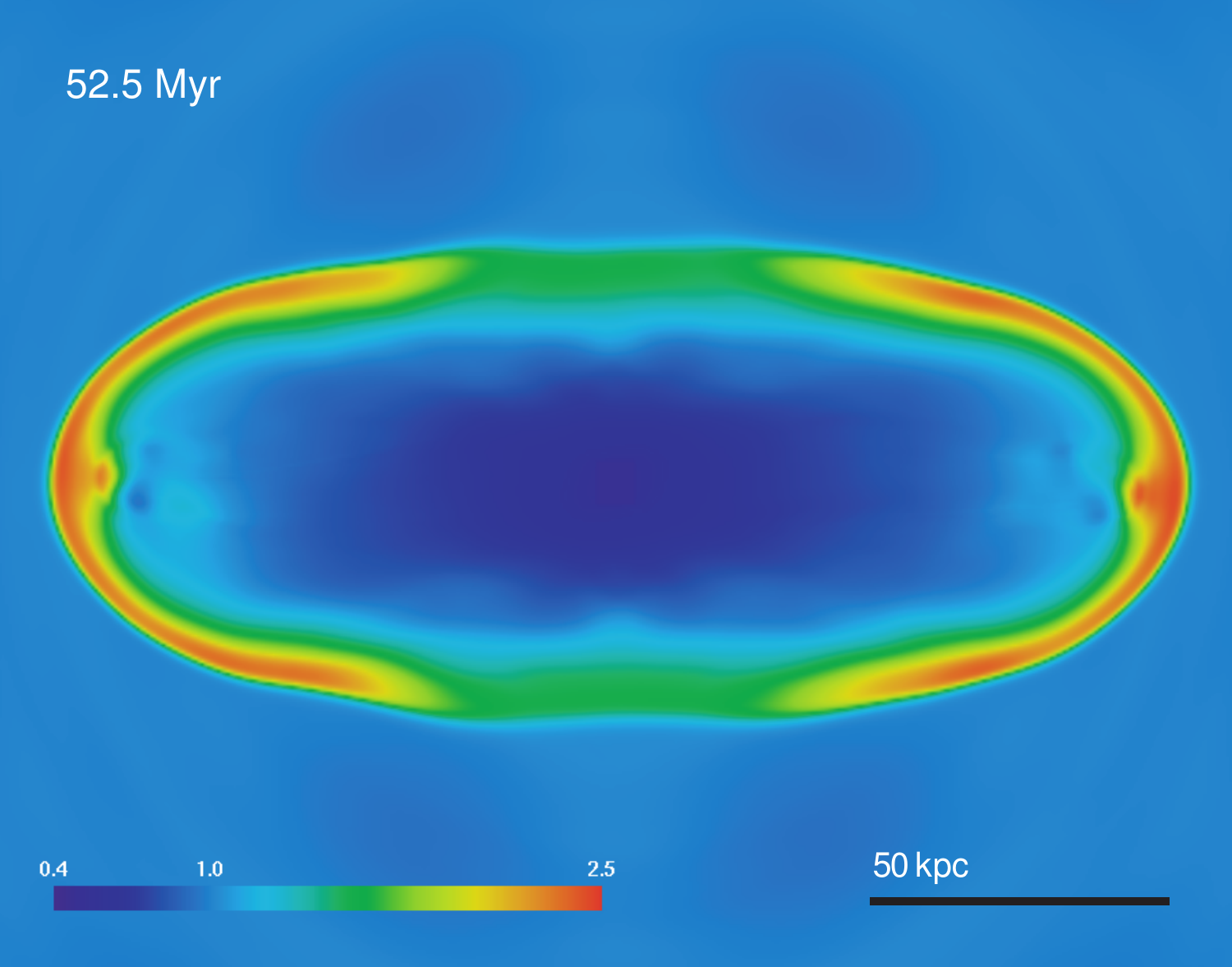}
  \caption{A slice through the midplane of the computational box showing the ratio of the pressure at 52.5 Myr into the RE simulation
to the initial conditions.  The pressure within 50 kpc of the cluster center has dropped by $\sim$20\% from the initial conditions.  The bow
shock is visible as an increase in pressure by a factor of $\sim$ 2.5 from the initial conditions.
(A color version of this figure is available in the
online journal.)}
\label{fig:REpressure_cut}
\end{figure}

The (azimuthally averaged) radial ICM pressure profile was calculated for each observation from the 
double $\beta$-profile model
temperature and density profiles just outlined, assuming an ideal gas equation
of state.  
Figures \ref{fig:REpressure} and \ref{fig:I13pressure} show example pressure profiles determined from the observed temperature
and density profiles 
along with azimuthally averaged ICM pressures
and AGN pressures extracted directly from the RE and I13 simulation.  Following the procedure used for density, only zones with $C_{jet}$ = 0 (pure ICM plasma)
were used to measure the ICM pressure profile.  Zones with $C_{jet} \geq$ 0.01 were used to measure the AGN pressure profile.
The top left panel of Figure \ref{fig:REpressure} shows the results of the 
double $\beta$-profile inversion (the ``observed'' profile) for the initial 
conditions.  Within $R_{C1}$ the observed profile underestimates the actual ICM profile by $\gtrsim$ 10\%.   This is due to the underestimate
of the electron density discussed in \S \ref{s:density}.  Outside of $R_{C1}$ the ICM pressure profile is measured with $\epsilon_{P} \le$ 10\%.
At the time the jets are turned off for the RE simulation, the top right panel of Figure \ref{fig:REpressure}, the ICM pressure is measured with
$\epsilon_{P}$ $<$ 20\% at all radii.  The signature of the bow shock in the ICM (solid line) can be seen at 35 kpc as a bump in excess over the
observed profile.  At 52.5 Myr, the lower left panel, the observed profile significantly overestimates the ICM pressure
by $\epsilon_{P}$ $>$ 10\% within $R_{C1}$.  The pressure $\sim$ 35 kpc behind the bow shock has dropped $\sim$ 20\% from the initial conditions
at this time as seen in Figure \ref{fig:REpressure_cut}.  By 157.5 Myr, the ICM has relaxed closer to equilibrium, and the observed profile
measures the actual pressure to $\epsilon_{P} \le$ 15\%.  

The observed pressure profiles for the I13 simulation shown in Figure \ref{fig:I13pressure}
reproduced the ICM profile to $\epsilon_{P} \le$ 45\% at all times, and at distances $\gtrsim 20-30$ kpc from cluster center
the observed profile was typically much better than that.  The intermittency of the jets in the I13 run produced a more complex
pressure distribution than the RE simulation. It cannot be captured by the smooth profile produced by the double $\beta-$profile inversion.  At
52.5 and 105 Myr, the actual pressure varies $\pm$30\% from the observed pressure within $R_{C2}$.  By 170.6 Myr, the strength of the bow shock
has diminished, and the error on the observed profile falls to $\epsilon_{P} \le$ 10\% outside $R_{C1}$.  Inside of $R_{C1}$, however, the effects
of the AGN activity produces a pressure structure poorly reproduced by the observed profile.  The measurement of the ICM pressure profile
from observation was reliable to within $\sim$ 20\% outside of regions strongly affected by jet related shocks.  Inside of shocked regions the 
measurement was only reliable to within $\sim$ 60\%.  This was true for both the RE and I13 simulations regardless of the observation orientation.

Following convention, the observed ICM pressure profiles were used to calculate X-ray cavity enthalpy on the
assumption that the ICM and cavity pressures were equal. 
The dotted lines in Figures \ref{fig:REpressure} and \ref{fig:I13pressure} show the average pressure in AGN plasma at each radial bin.
This pressure could only be observationally measured if the cavity were in exact pressure balance with the ICM and the exact cluster-centric
distance of the cavity was known.  Here we discuss how closely observation matches the AGN plasma profile assuming the cavity location is known.
\S \ref{s:cav_enthalpy} discusses the effect of projection on inferred pressure.

The AGN plasma pressure in the RE simulation, as shown in 
Figure \ref{fig:REpressure}, roughly follows the ICM pressure at 26.3 and 157.5 Myr
except for high pressure at the ends of the jets where momentum
flows drive the cavities outward (OJ10).  At 52.5 Myr, the AGN plasma pressure differs by as much as a factor of three from the actual 
ICM pressure from 30-100 kpc. 
This discrepancy can be explained by the influence of the bow shock.  Referring to Figure \ref{fig:REpressure_cut}, shocked ICM material between 
projected distances of 30-100 kpc raised the average ICM pressure over the lower pressure inside of the jet cocoon.  The observed pressure profile, which
is sensitive to the ICM pressure, is
$\sim$ 75\% greater than the AGN plasma pressure within $R_{C1}$ at this time.

For the I13 simulation at 26.3 Myr in Figure
\ref{fig:I13pressure}, there is a significant difference between the AGN and ICM profiles from 25-35 kpc also due to the effects 
of the bow shock.  At this time the observed pressure profile overestimates the AGN plasma pressure by
$\sim$ 50\% within 20 kpc.  At 52.5 Myr, the AGN and ICM profiles approximately agree with the exception at the ends of the jets.  Here the
observed profile matches the actual AGN plasma pressure to within 13\%.  The observed, AGN, and ICM profiles all agree at 52.5 Myr to within $\sim$ 30\%
except the ends of the jets.  The intermittency of the jets impacts how well the observed profile reproduces the AGN pressure.  At 105 Myr, when
the jets are inactive, the observed pressure overestimates the AGN pressure within 30 kpc while at 170.6 Myr, when the jets are active, it underestimates it. 
Inferring AGN plasma pressure from observational measurement of the ICM pressure at a specific radius was reliable only to $\sim$ 75\% from the 
RE and I13 observations.

\subsubsection{Cavity System Volume}
\label{s:volume}

\begin{figure}
  \includegraphics[height=.8\textheight]{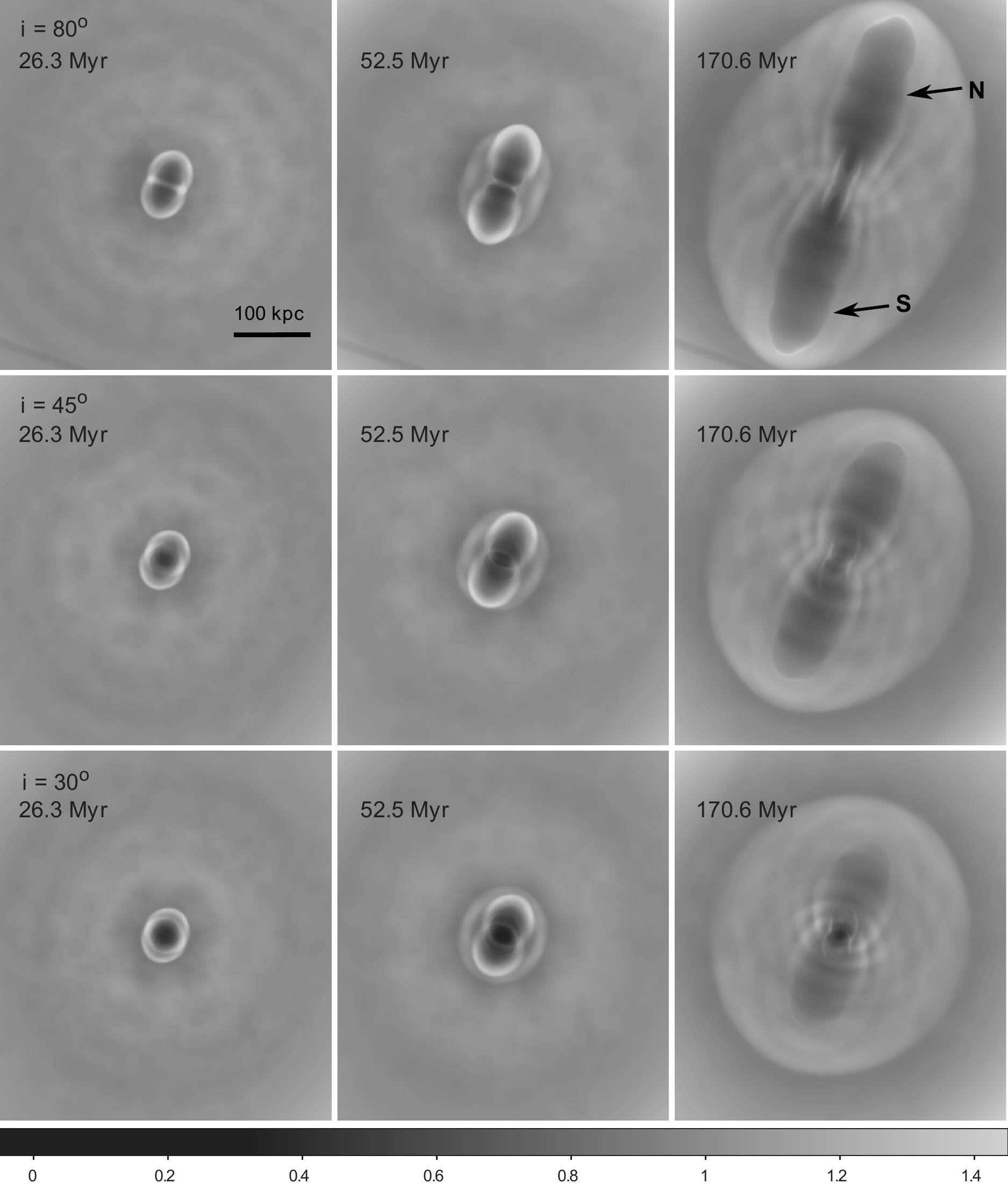}
  \caption{Synthetic observations as in Figure \ref{fig:REobs} of the I13 simulation.}
\label{fig:I13obs}
\end{figure}

\begin{figure}
  \includegraphics[height=.5\textheight]{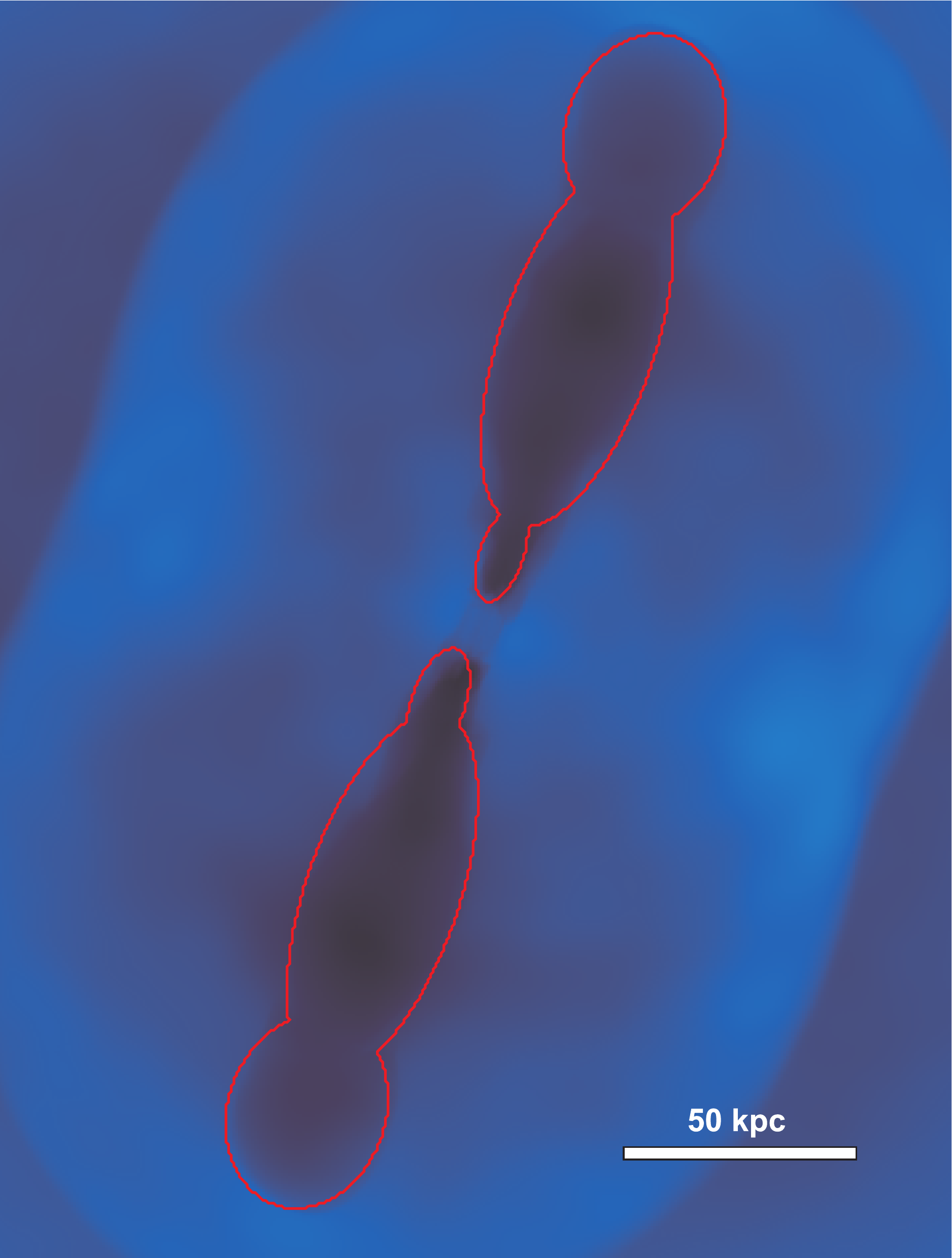}
  \caption{Example of area enclosed by ellipses fit by eye (\emph{solid line}) to a 1.5-2.5 keV observation of the RE simulation at 131.3 Myr divided
by a best-fit double $\beta-$profile.
(A color version of this figure is available in the
online journal.)}
\label{fig:ellipse}
\end{figure}

Cavity volumes were estimated from 1.5-2.5 keV observations at each analyzed epoch 
for both RE and I13 simulations.  As already noted, each cavity 
was fit by eye with a set of ellipses \citep{birzan04}.  
For each projection angle, $i = 80^{\degr}, 45^{\degr},~{\rm and}~ 30^{\degr}$, 
the cavities were assumed in this measurement to be in the plane of 
the sky in order to test the effects of an inaccurate or unknown value for the 
inclination.  These observed cavity volumes were calculated assuming them to be ellipsoids
of revolution around the major axis of each projected ellipse.
By using multiple ellipses to cover each 
projected cavity we were better able to define the outer edge of the X-ray cavity.  Figure \ref{fig:ellipse} shows an example of the area enclosed 
by the ellipses
chosen for the RE simulation at 131.3 Myr observed at $i$ = 80$^{\text{o}}$.
In general, it was difficult to define the edge 
of the cavities for the RE simulation once the cavities extended past $R_{C1}$. For the I13 simulation the cavities were often 
outlined with a bright rim (see Figure \ref{fig:I13obs}), making the edge (taken to be the inside of the rim) easier to find.

The actual cavity volumes were computed by integrating the volume in the
simulation data with $C_{jet} \geq 0.01$ (partially AGN plasma).
Observed and actual volumes are compared in Figure \ref{fig:volume}. Since we expect
a projection bias due to foreshortening along the jet axis (see Appendix \ref{a:ellipse}), we plot 
the observed volume divided by $a_{p} / a$, where $a_{p}$ is given by Equation \ref{eq:ellipsecorr} and $a$ is the actual length of a single best fit ellipse,
normalized by the actual volume.  The scatter in the measurements without correcting for this projection bias was $\sim$ 50\%.  
Two features stand out in the comparisons in Figure \ref{fig:volume}. First, the $a_{p} / a$ correction reduced the scatter due to the projection bias
to approximately 10-15\% for both RE and I13.  The second obvious feature of the comparison is that the observed cavity volume estimates
tend to be modestly smaller than the actual volumes for RE but not for I13.
The reason for this has to due with the different shapes of the cavities between RE and I13.  I13 retains a nearly elliptical area
at all times while RE developed a non-elliptical shape (see Figures \ref{fig:REobs} and \ref{fig:I13obs}), which required many ellipses to fit.
Fitting ellipses will tend to underestimate a non-elliptical shape if the observer requires that none of the fits extend beyond the cavity edge.
Despite this limitation, and the subjective, observer-dependent nature of
the process, our observed volume estimates generally agree with the actual volumes to within about $\pm 50$\% (omitting the $a_{p} / a$
correction).

The observations used to determine cavity analysis did not include any noise representing X-ray counts or intrumental effects.  Low counts at large
cluster centered distances would make cavity edges in those regions more difficult to identify.  The long axis of all of the observed cavities, 
roughly aligned with the jet, would likely be underestimated given these conditions, which would reduce the measured volume by the same amount.

\subsubsection{Cavity System Enthalpy}
\label{s:cav_enthalpy}

The observational estimates for the ICM pressure profile and cavity system 
volume presented above were used to derive the total cavity enthalpy, $H_{obs}$.
The cavity volume enclosed by the chosen ellipses for each observation was discretized into 1 kpc$^{3}$ volume elements corresponding to the 1 arc sec resolution
of the images for a cluster distance, $D_{L}$ = 240 Mpc.  The pressure in 
each cavity volume was determined from the observed pressure profile and the 
projected cluster-centric distance of the volume elements.  The total enthalpy in the cavity system for each observation was given by
\begin{align}
H_{obs} = \frac{5}{2}\displaystyle\sum_{i}^{N_{elements}}P_{i}\left(1\,kpc^{3}\right).
\end{align}

\begin{figure}
  \includegraphics[height=.3\textheight]{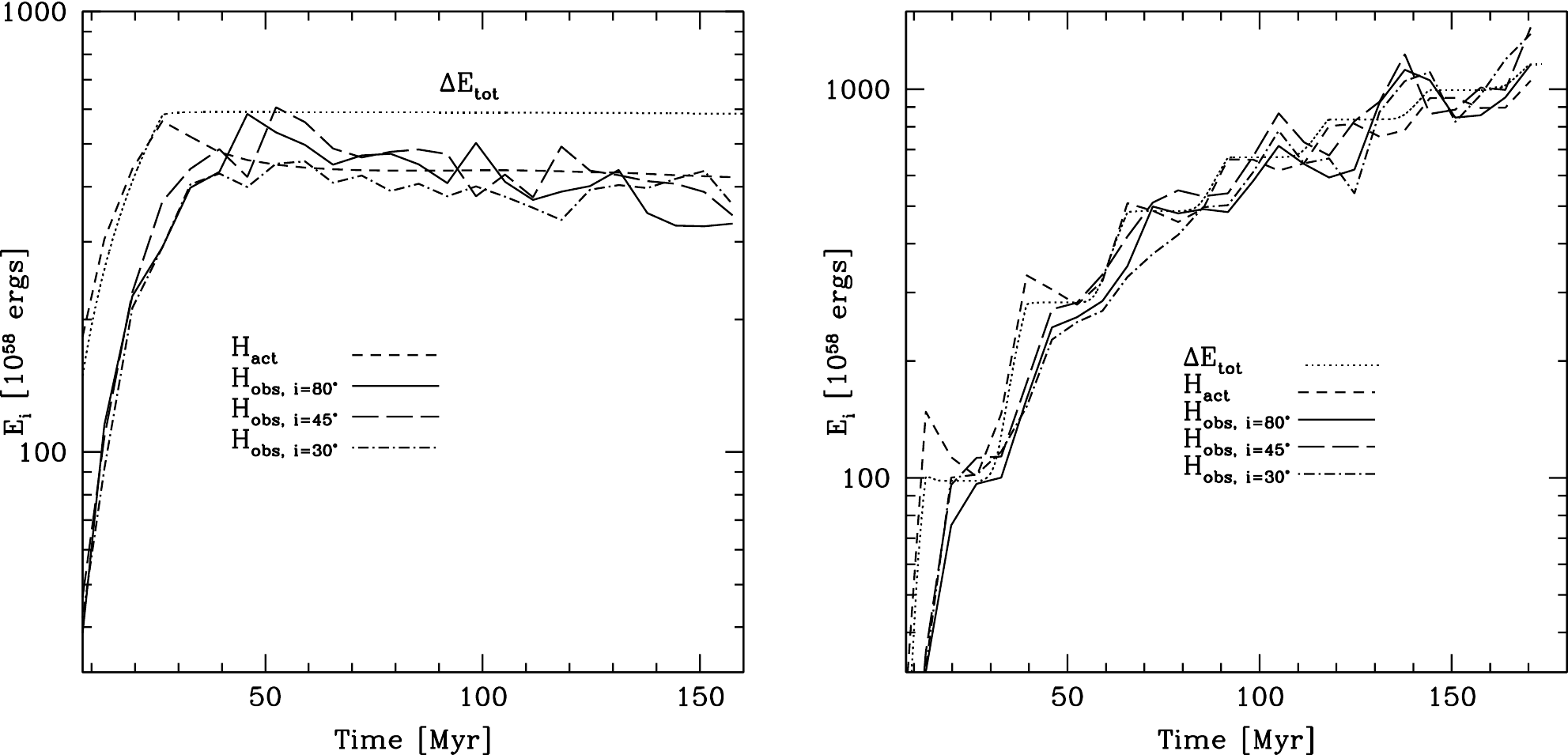}
  \caption{Comparison between the actual enthalpy in AGN plasma, $H_{\text{act}}$, and observed enthalpy, $H_{\text{obs}}$, for the RE simulation 
(\emph{left}) and the I13 simulation (\emph{right}).  H$_{\text{obs}}$ measured from observations at three inclination angles are shown.   
The total energy added to the computational grid by the AGN, $\Delta E_{tot}$, is shown as a dotted line for reference.}
\label{fig:enthalpy}
\end{figure}

Figure \ref{fig:enthalpy} shows comparisons of the total energy
added to the simulation volumes by jets, $\Delta E_{tot}$, the actual enthalpy in the cavity systems, $H_{act}$, 
and the above observed cavity enthalpy estimates, $H_{obs}$, throughout 
the RE and I13 simulations at three inclinations.  The value for $H_{act}$ was computed from the pressure and total volume of each voxel in the
simulation data with $C_{jet} \geq 0.01$ to be consistent with the synthetic observations (see \S \ref{s:synobs}).  This conservative cut meant 
that some ICM enthalpy was included in $H_{act}$ values, making it possible for $H_{act} > \Delta E_{tot}$.

Several features stand out in comparing the various energy measures.  The first feature is that $H_{obs}$, $H_{act}$, and $\Delta E_{tot}$
all agree with each other within about a factor of two for a given simulation and inclination angle.  
Second, the comparisons of enthalpy, both observed and actual, and the total energy measures is
better for the intermittent jet simulation, I13, than for the terminated, RE, case.
This is consistent with the analysis of the 
simulations reported in OJ10. In particular, they
noted that about 50\% of the jet energy ($\Delta E_{tot}$) injected during the
active phase of either the I13 or RE model had been converted into ICM thermal or
kinetic energy by the end of the simulation.  The remaining energy increment was mostly
gravitational potential energy in the ICM or thermal energy in the jet cocoon\footnote{Relatively smaller energy
increments are contained at a given time in jet kinetic energy and magnetic fields.},
which is roughly what $H_{obs}$ measures.  Approximately 30\% of $\Delta E_{tot}$ ended up as gravitational potential energy in the ICM 
for the RE simulation.  By contrast, a much smaller fraction, $\sim 15$\%, of $\Delta E_{tot}$
in the intermittent jet, I13, simulation is converted to ICM gravitational
by the simulation's end.  Thus, we should expect a closer match between $\Delta E_{tot}$ and $H_{act}$ in that case. 

Another striking feature for both panels of Figure \ref{fig:enthalpy} is the consistency in $H_{obs}$
among the inclination angles for a given simulation and epoch.  This is due to two competing projections effects. In particular, the estimated volume generally decreases 
with decreasing inclination
angle as discussed in \S \ref{s:volume}, while the projected distance from cluster center decreases, making a cavity appear to be in a higher pressure 
environment then 
it actually is.  To see the net effect refer to Figures \ref{fig:REpressure} and \ref{fig:I13pressure}. It is evident that the ICM pressure profile 
can be approximated as a power law, $P = P_{0}(r / r_{0})^{\alpha}$, over
the projected distances the cavities occupy (40-200 kpc) for most of the simulated time.  Then, assuming a given cavity is a cylinder with radius $R$ 
extending from a projected distance $r_{1}\,sin\,i$ to $r_{2}\,sin\,i$ (see Appendix \ref{a:ellipse} regarding the use of $sin\,i$) and 
estimating $\alpha \approx$ -1 we would predict that the enthalpy $H$ would be
\begin{align}
H &= \pi\frac{5}{2}\,R^{2}\,p_{0} \int_{r_{1}\,sin\,i}^{r_{2}\,sin\,i} \left( \frac{r}{r_{0}} \right)^{-1}\,dr \\
  &= \pi\frac{5}{2}\,R^{2}\,p_{0}ln \left(\frac{r_{2}}{r_{1}} \right),
\end{align}
which is independent of $i$.

For the RE simulation on the left panel of Figure \ref{fig:enthalpy} $H_{obs}$ always underestimated
$H_{act}$ while the jets were active.  Recall that $H_{obs}$ depends on the
observed estimate of the ICM pressure distribution and, from \S 
\ref{s:pressure}, that the observed pressure profile underestimated the AGN plasma pressure (the cavity
pressure) while the jets were active. In short, during those times
the cavities are over-pressured as they drive moderate strength shocks into the
ICM. This is consistent with comparisons shown in Figure \ref{fig:REpressure}. Further into the simulation we see $H_{act}$ declining.  The cavities
are rising buoyantly in the cluster while maintaining approximate pressure equilibrium (see OJ10).  The thermal energy in the cavities dropped as this
energy was transferred to gravitational potential energy.  The observed values follow this
trend, remaining within $\approx$ 50\% $H_{act}$.

The I13 simulation on the right panel of Figure \ref{fig:enthalpy} shows a different evolution of $H_{act}$ and $H_{obs}$ because of the different AGN history.  
There is a step-like growth of $H_{act}$ due to the intermittency of the jets.  The time delay between 
the peaks of $H_{act}$ 
and $H_{obs}$ at each step is due to the difference in evolution of the observed ICM pressure and the actual AGN plasma profiles.  When the jets turn on the
cavities are over-pressured with respect to the ICM, but they eventually expand to approximate pressure balance.  Prior to the 
expansion of the cavities, however, the higher energy content of the cavities cannot be accurately measured by the procedure described in \S \ref{s:pressure}.  
Therefore, an increase of $H_{obs}$ lags behind an increase of $H_{act}$.  

\subsection{Ages}
\label{s:ages}

Three characteristic timescales are commonly employed for determining cavity age.  For a cavity centered at a projected distance $r_{p}$ 
from cluster center,
radius $R$, cross section $S$, drag coefficient $C$, and volume $V$ these times are: 1) the buoyant rise time $t_{buoy} \approx
r_{p}\sqrt{CS/(2gV)}$, 2) the ``refill time'' $t_{r} = 2\sqrt{R/g}$, and 3) the sound crossing time $t_{c} = r_{p}/c_{s}$ \citep{birzan04}.  These times
can be compared to known ages from the simulations given measurements of the sound speed, $c_{s}$, and $g$ from the synthetic
observations.  Measurements of each age were made from observations of the \textbf{N} and \textbf{S} cavities (see Figures \ref{fig:REobs} and 
\ref{fig:I13obs}) at the end of 
the RE simulation, $t = 157.5$ Myr, and I13 simulation, $t = 170.6$ Myr, at three different inclination angles.  In these observations we represented
each cavity as a single ellipsoid with semi-major axis $a$ and semi-minor axis $b$.
Following a procedure similar to \citet{birzan04}, the value of $r_{p}$ was the projected distance
from cluster center to the center of the cavity, the radius was given by $R = \sqrt{ab}$, and the cross section was given by $S = \pi\,b_{max}^{2}$, 
where $b_{max}$ was half the maximum azimuthal width of the cavity.  The volume $V$ was determined for 
each cavity following the method described in \S \ref{s:volume}.  The sound speed was given by $c_{s} = \sqrt{\gamma k<T_{ICM}>/(\mu m_{H})}$.  
At the end of the RE simulation $<T_{ICM}> =$ 2.77 keV giving $c_{s}$ = 941 km s$^{-1}$, and at the end of the I13 simulation $<T_{ICM}> =$ 2.82 keV, 
giving $c_{s}$ = 950 km s$^{-1}$.  Refer to Appendix \ref{a:gravity} for details on estimating $g$, which we assumed to be constant, from the synthetic 
observations.  We let $C$ = 1 for simplicity.

Table \ref{tab:age} shows the measured parameters and age estimates for each observation at the end of the RE and I13 simulations.  
Both $t_{buoy}$ and $t_{c}$ are affected by projection.  For this reason, we would expect the 
buoyant rise time to vary as $t_{buoy} \propto r_{p}/\sqrt{a} \propto \sqrt{sin\,i}$ due to projection effects on both the projected distance 
and semi-major axis of the cavity.  The sound crossing time, however,
should vary more rapidly with inclination as $t_{c} \propto r_{p} \propto sin\,i$.  The cavity pairs for both simulations approximately show
this trend for $t_{buoy}$ and $t_{c}$.  The refill time shows a weaker dependence on inclination angle because $t_{r} \propto \sqrt{(ab)^{1/2}} 
\propto (sin\,i)^{1/4}$ (recall that we assumed constant $g$ in calculating ages).  
An important aspect of Table \ref{tab:age} was that for most cases $t_{buoy} < t_{c}$, 
which implies a terminal buoyant velocity greater than the sound speed.  
This unphysical result could have been avoided in a number of ways.
In their analysis of Hydra A, for example, \citet{wise07}
represented the cavity system as a series of spherical bubbles.  Approximating the end of the \textbf{N} cavity of the RE simulation observed at 
$i$ = 80$^{\text{o}}$
as a sphere would increase $r_{p}$ to $\sim$ 200 kpc and would decrease $V$ to $(4/3)\pi\,r^{3} \sim 1.4\times10^{5}$ kpc$^{3}$.  Given these measurements
for an outer spherical cavity, $t_{buoy} \sim$ 244 Myr while $t_{c} \sim$ 208 Myr.  Another approach may have been to assume $C >$ 1 similar to 
values empirically estimated by \citet{jones05}.  
The parameters and measurements used in a buoyant rise model are not very well constrained.  A buoyant model also did not properly capture the evolution
of the cavities in the RE or I13 simulations (see OJ10).  For these reasons we chose not to use $t_{buoy}$ as the cavity age.
The equations for $t_{r}$ and $t_{c}$ are related, and the models only differ in the length over which material moves.  By convention, $t_{r}$ uses the cavity
radius, which is not affected by projection.  We instead use the values of $t_{c}$ for the cavity ages 
in subsequent calculations so that our analysis demonstrates the dependence on projection.
When projection did not greatly effect our measurements $t_{c}$ was reliable to within $\pm 20$\%.

\subsection{Cavity Power}
\label{s:lum}

The enthalpy and age of a cavity system are typically combined into a characteristic
power called the cavity power $P_{cav} = H / t$.  If $P_{cav}$ can be deposited into 
the ICM it may balance the cooling in the host cluster.  It is therefore common practice to compare $P_{cav}$ with the cooling 
\citep[\emph{e.g.,}][]{birzan04, rafferty06, cavagnolo10}.  
The cavity power should represent a lower limit to the total luminosity of the AGN because it does not account for energy already deposited into 
the ICM or other forms
of AGN plasma energy such as magnetic, potential, or kinetic (see \S \ref{s:cav_enthalpy}).  Table \ref{tab:lum} shows a comparison of the observed 
$P_{cav,obs}$ with the actual
total average jet luminosity, $L_{jet,act}$, for both RE and I13 observed at the three different 
inclinations at the end of each simulation.  $P_{cav,obs}$ was computed as $H_{obs} / \langle t \rangle$, where $\langle t \rangle$ was the average of the ages 
in Table \ref{tab:age} for the \textbf{N} and \textbf{S} cavities, while the
mean jet power, $L_{jet,act} = \Delta E(t) / t$, where $t$ = 157.5 Myr and $t$ = 170.6 Myr for the RE 
and I13 simulations respectively.  The differences between $P_{cav,obs}$ and $L_{jet,sim}$ at these times were characteristic of early times.

Observations of the RE and I13 simulations produced $P_{cav,obs}$ increasing with decreasing inclination primarily due to projection effects on 
measuring the ages as discussed in \S \ref{s:ages}.  For the RE simulation, $P_{cav,obs} < L_{jet,act}$ as we would expect for all but the smallest inclination
angles.  The I13 observations, however, resulted in $P_{cav,obs} > L_{jet,act}$ at all orientations.  The underestimate of cavity system age was the dominant 
reason $P_{cav,obs}$ exceeded $L_{jet,act}$ in these observations.
For measurements from observations at $i$ = 80$^{\text{o}}$, when the
error on the assumed inclination is small, $P_{cav,obs}$ was within $\sim$ 40\% of $L_{jet,act}$, and $P_{cav,obs}$ was within
a factor of three of $L_{jet,act}$ across all of the observed inclination angles.


\section{Conclusions}
\label{s:conclusion}

We have presented an analysis of the reliability of common techniques used to extract X-ray cavity enthalpy, age, and mechanical
luminosity from X-ray observations of cavity systems.  By utilizing synthetic X-ray observations of detailed simulations we were able to
directly compare observationally determined and actual values from the simulations.  The important results from this work are:

$\bullet$ The synthetic observations of the I13 simulation show bright rims outlining the cavities at each analyzed epoch out to 170.6 Myr while
the RE simulation does not show bright rims.  The difference in the AGN history represented by each model accounts for this difference.  I13 had
periodic injection of energy into the cavities throughout the simulation while RE deposited all of its energy early on.

$\bullet$ Observationally measuring X-ray cavity enthalpy is reliable to within approximately a factor of two across a wide range of age and inclinations
for the models of jet intermittency presented here.  Several steps go into determining the enthalpy, and each may introduce significant
errors.  Extracting the ICM electron density profile, for example, was reliable to within $\sim$ 20\% outside of regions strongly influenced by shocks.
Inside recently shock influenced regions the error was $\sim$ 40\%.  Combining this and the temperature profile into the pressure inside of the cavity at a 
given cluster-centric distance may not be as accurate.  During
periods of jet activity, the observationally determined pressure may differ by as much as $\sim$ 75\% from the cavity pressure.  This is related to the 
supersonic speeds of the jets through the ICM and the consequent post-shock pressure enhancement.  Our measurements of cavity volume were within $\pm$50\% 
of the actual total cavity system volume.  
This process is subjective, however, and a more robust and objective method for finding and outlining cavities should be developed.  The overall 
effect of each of these measurements is contained in the factor of two reliability of enthalpy measurements.
The energy required to offset cooling in clusters can be characterized as $\eta PV$.  An approximate factor of two span in $\eta$ in our tests is due
largely to uncertainties in the measurement of $PV$.

$\bullet$ The determination of cavity age from one or more of the commonly used age estimates could potentially be misleading.  The buoyant rise
model was not an accurate description of the evolution of the cavities in our simulations.  A simple application of this model implied unrealistic
terminal velocities greater than the sound speed.  The refill time model produced ages within $\pm$15\%
of the correct age regardless of the error on assumed inclination.  It relied on an accurate measurement of the gravitational acceleration,
however, which assumes the cluster to be in hydrostatic equilibrium.  This assumption may be not be valid for a given cluster.  We preferred to use the
sound crossing time as a simple and fairly robust model for cavity age.  For a well constrained inclination angle, our measurements were within $\pm$20\% of the
actual cavity age.

$\bullet$ Observationally measuring the cavity power produced values within a factor of $\sim$ 3 of the average total jet luminosity from our 
simulations regardless of assumed inclination angle.  
The observed cavity power was within 40\% of the jet luminosity if the projection effects were negligible.  At all observed inclination angles for the
I13 simulation and $i$ = 30$^{\text{o}}$ for the RE simulation the cavity power overestimates the average jet luminosity largely due to underestimates
in the cavity system age.

\begin{acknowledgments}
This work was supported at the University of Minnesota by NSF grant AST0908668
and by the University of Minnesota Supercomputing Institute. PJM and SMO were supported in 
part by the Graduate Dissertation 
Fellowship at the University of Minnesota.  SMO was also supported by NASA Astrophysics Theory Program Grant NNX09AG02G.  We
are grateful to Brian McNamara and Paul Nulsen for very fruitful
conversations and to an anonymous referee for help in improving the
original manuscript.  
\end{acknowledgments}

\appendix

\section{Fitting $\beta$ Profiles}
\label{a:fitting}
To calculate a best fit to the brightness distributions comparable with observations, the flux from the synthetic observation was scaled by the net counts 
from Chandra observations of Hydra.  Counts were taken from the $evt2$ files in the Chandra archive (ObsIDs 575 and 576, 
Chandra calibration program; ObsIDs 4969 and 4970, program 05800556, P.I. McNamara) \citep{wise07}.  The total number of counts with 
energies from 1.5-2.5 keV in a circular region centered on the cluster with a diameter equal to the size of the synthetic observations, 
12.3 arc min a side, was extracted for each observation.  A total of $N_{c} = 451345$ counts for the four observations were detected in this 
region over 227 ksec.  The scaling factor $\xi$ was calculated as
\begin{align}
\xi = \frac{N_{c}\,\,counts}{\displaystyle\sum_{i}^{N_{pixels}}S_{i}\,\,erg\,s^{-1}\,cm^{-2}},
\end{align}
where $S_{i}$ is the flux for a given pixel observed over a band from 1.5 to 2.5 keV.  $\chi_{norm}^{2}$ was then calculated from the simulated average counts 
in a bin, the product of $\xi \langle S_{i} \rangle$, assuming Poisson statistics.  The downhill simplex method was used to minimize $\chi_{norm}^{2}$
by simultaneously fitting $S_{0i}$, $R_{C,i}$, and $\beta_{i}$.  Both the RE and I13 grids extended only 300 kpc.  To ensure that our fits were not affected 
by the truncation of the atmosphere, additional azimuthally averaged data points were generated from a synthetic observation of the cluster discussed in 
\S \ref{s:cluster} extending the initial conditions to 600 kpc.

\section{Projected Semi-major Axis of an Ellipsoid}
\label{a:ellipse}
\begin{figure}
  \includegraphics[height=.38\textheight]{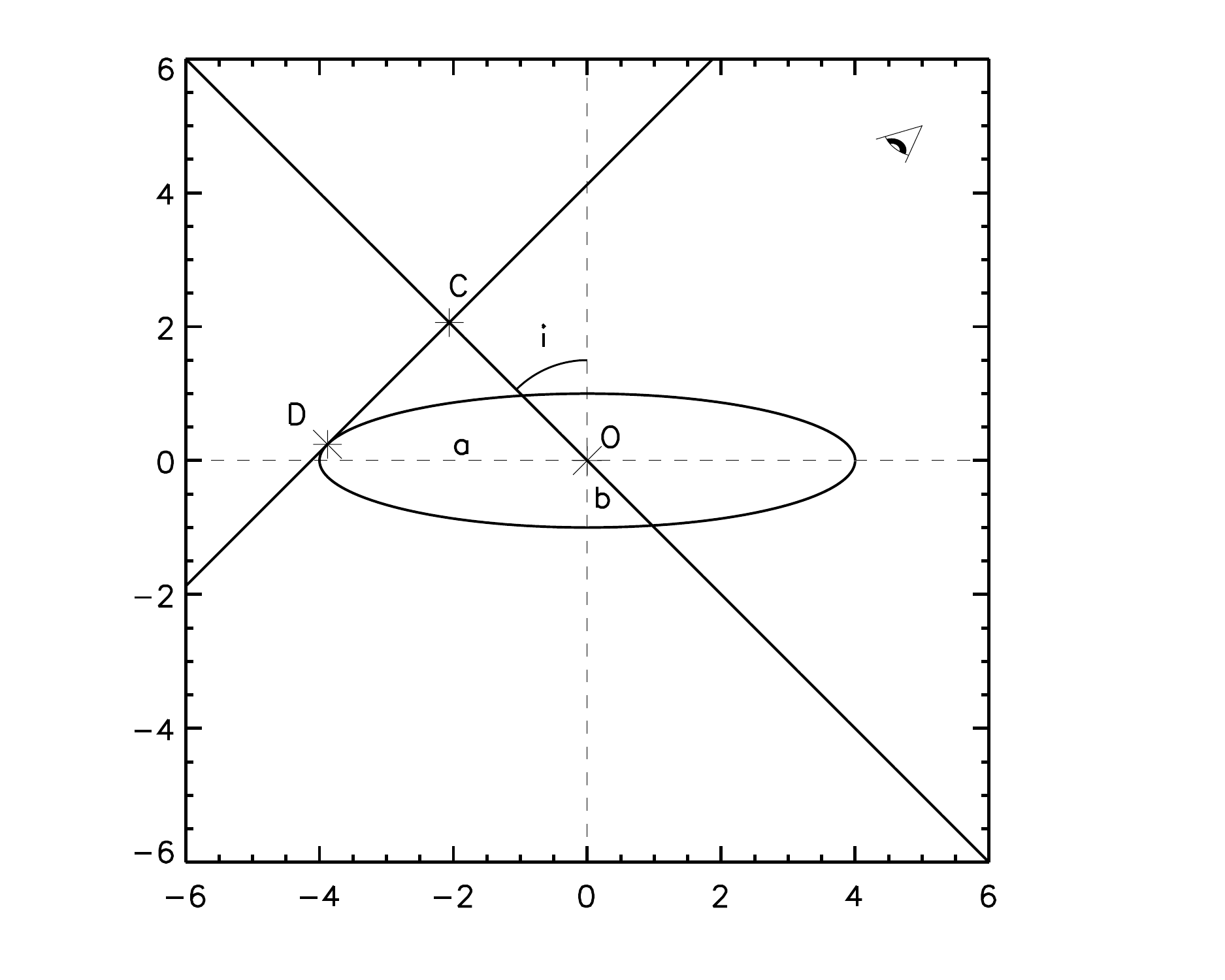}
  \caption{Diagram showing an ellipse with an aspect ratio of 4 observed at an inclination angle $i = 45^{\degr}$.  The length of the line segment from 
point $\bar{OC}$ is the projected semi-major axis.  The segment is defined by a tangent line and the orthogonal line passing through the origin.}
\label{fig:projection}
\end{figure}

The projected semi-major axis of revolution of an ellipsoid can be determined from the inclination angle, $i$, and the un-projected aspect ratio taken to 
be $\psi = a/b$, where
$a$ is the semi-major axis and $b$ is the semi-minor axis along the other two dimensions.  Figure \ref{fig:projection} shows an example of the geometry of the 
problem where the line segment $OC$ is the projected semi-major axis.  We start with an ellipse defined by
\begin{align}
\frac{x^2}{a^2} + \frac{y^2}{b^2} = 1
\end{align}
lying in a plane that includes the line of sight with a tangent line of slope
\begin{align}
\frac{dy}{dx} = -\frac{a^2}{b^2}\frac{y}{x} = -\psi^{2}\frac{y}{x} = tan\,i
\end{align}
from the observer.
The coordinates $D = \{x_{0},y_{0}\}$, where the tangent line intersects the ellipse, are given by
\begin{align}
x_{0} = -\left(\frac{1}{\psi^{4}tan^{2}i\,b^2} + \frac{1}{a^2}\right)^{-1/2},
y_{0} = \frac{1}{\psi^2tan\,i}x_{0}.
\end{align}
The equation for the tangent line is then
\begin{align}
y = y_{0} + tan\,i(x - x_{0}).
\label{eq:ellipse1}
\end{align}
This line will intersect the orthogonal line passing through the origin given by
\begin{align}
y = -\frac{x}{tan\,i}.
\label{eq:ellipse2}
\end{align}
Solving for $x$ and $y$ where Equations \ref{eq:ellipse1} and \ref{eq:ellipse2} intersect one finds the length $\bar{OC}$ to be
\begin{align}
a_{p} \equiv \bar{OC} = \left(\frac{\left(\frac{1}{\psi^2} + tan^{2}i\right)^2\left(1 + tan^{-2}i\right)}{\left(\frac{1}{\psi^{4}tan^{2}i\,b^2} + a^{-2}\right)
\left(1 + tan^{2}i\right)}\right)^{1/2}.
\label{eq:ellipsecorr}
\end{align}

\begin{figure}
  \includegraphics[height=.38\textheight]{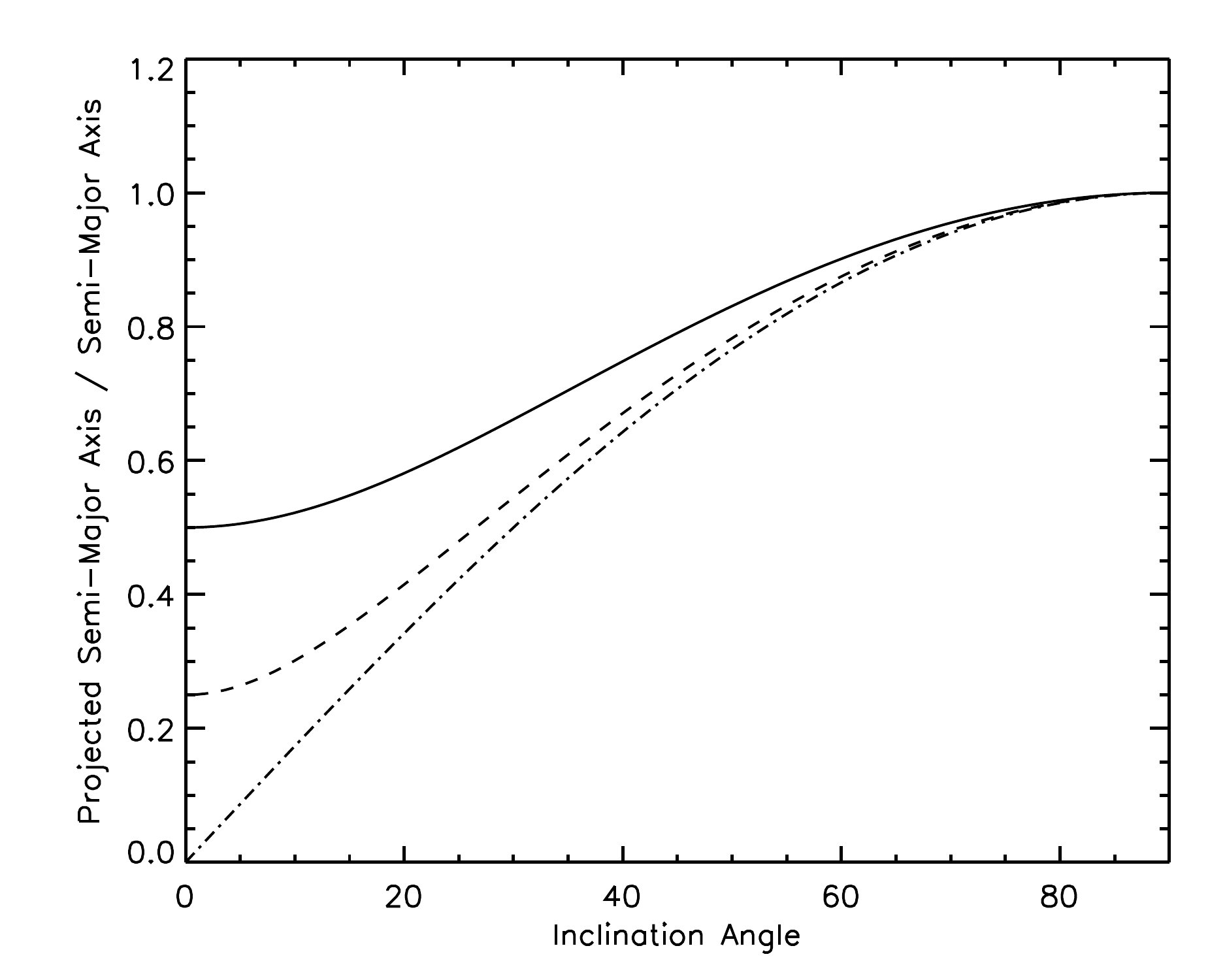}
  \caption{The projected semi-major axis of an ellipse with an aspect ratio of 2 (\emph{solid curve}) and aspect ratio of 4 (\emph{dashed curve}) 
as a function of inclination angle.  A $sin\,i$ (\emph{dash-dot curve}) curve is shown for comparison.  At large inclination angles a simple $sin$ 
curve closely approximates the projected semi-major axis for both aspect ratios.}
\label{fig:projection_corr}
\end{figure}

Figure \ref{fig:projection_corr} shows the projected semi-major axis normalized by the actual semi-major axis, $a_{p} / a$, as a function of $i$ 
for an $\psi =$ 2 and $\psi = $ 4.  At large inclination angles ($i \ge 45^{\text{o}}$) $sin\,i$ approximates $a_{p} / a$ to within 12\%.
For simplicity we typically chose to correct an observed semi-major axis in that regime with $sin\,i$, as it does not require that $\psi$ be known.

\section{Gravitational Acceleration}
\label{a:gravity}
The gravitational acceleration at a given projected cluster radius $r_{p}$ can be determined from the synthetic observations using an 
``observed'' measurement of the distribution of gravitating mass.
An estimate of the mass within a radius $r_{p}$ from the synthetic observations was be made assuming the ICM plasma was in hydrostatic equilibrium with an 
electron density profile derived from a double $\beta$-profile (Equation \ref{f:betaden}), and dominated by the pressure of the thermal gas.  
>From these assumptions the distribution, given by \citet{xue00}, is
\begin{align}
M(r_{p}) = \displaystyle\sum_{i,j}M_{i}(r_{p})\left(\frac{n_{e,i}(r_{p})}{n_{e}(0)}\right)\left[1 + \left(1 - \frac{\beta_{j}}{\beta_{i}}\frac{r_{p}^{2} 
+ R_{C,i}^{2}}{r_{p}^{2} + R_{C,j}^{2}}\right)\frac{\tilde{n_{e,j}}}{\tilde{n_{e,i}} + \tilde{n_{e,j}}}\right]
\end{align}
where
\begin{align}
M_{i}(r_{p}) = 3\beta_{i}\gamma\frac{T_{ICM,i}(0)r_{p}}{G\mu m_{p}}\frac{r_{p}^{2}}{r_{p}^{2} + 
R_{C,i}^{2}}\left(\frac{n_{e,i}(r_{p})}{n_{e,i}(0)}\right)^{\gamma - 1}
\end{align}
and
\begin{align}
\tilde{n_{e,i}} = n_{e,i}(0)\left(1 + \frac{r_{p}^2}{R_{C,i}^{2}}\right)^{-3\beta_{i}}.
\end{align}
The gravitational acceleration, $g$, was then computed as $g(r_{p}) = GM(r_{p})/r_{p}^2$. 
The resulting value of $g$ varied by approximately a factor of two over most of the observed cluster (10-300 kpc).  In deriving the timescales discussed in 
\S \ref{s:ages} a single value for $g$ is required.  We adopted the value of $g$ to be a simple average over the observed $g(r_{p})$ profile.  This value for $g$
was within a factor two of the actual gravitational acceleration used in the simulation, $g_{act}(r_{p})$ (Equation \ref{f:grav}), over the range 10 kpc $\le r_{p} \le$ 300 kpc.




\begin{figure}
  \includegraphics[height=.3\textheight]{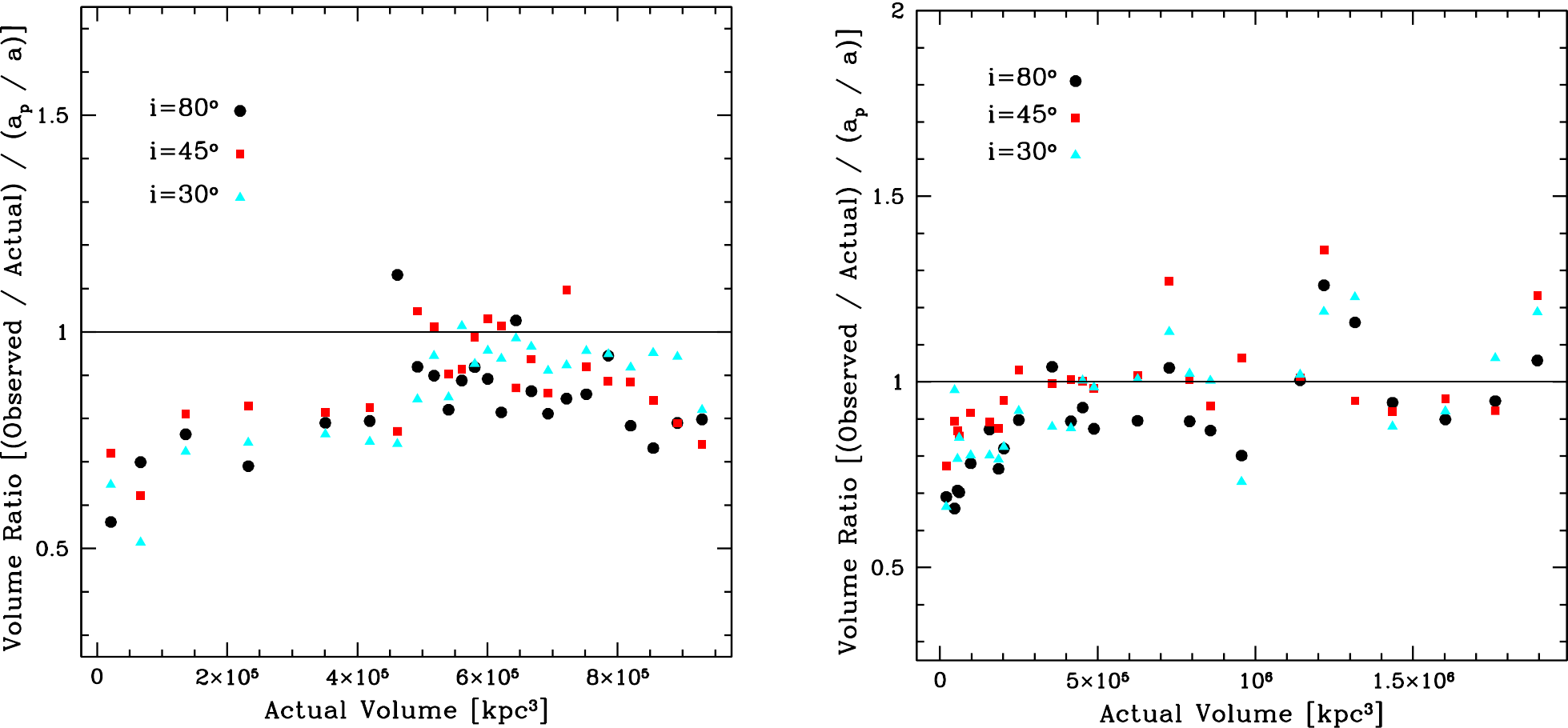}
  \caption{The observationally determined total cavity volume to the actual total cavity volumes from each every analyzed epoch for the RE 
simulation (\emph{left}) and the I13 simulation (\emph{right}).
(A color version of this figure is available in the
online journal.)}
\label{fig:volume}
\end{figure}

\begin{deluxetable}{cccccccccccc}
\tabletypesize{\scriptsize}
\tablecolumns{12}
\tablewidth{0pt}
\tablecaption{Cavity Ages\label{tab:age}}
\tablehead{
    \colhead{Model}
  & \colhead{$i$}
  & \colhead{Cavity}
  & \colhead{$r_{p}$}
  & \colhead{$a$}
  & \colhead{$b$}
  & \colhead{$b_{max}$}
  & \colhead{$V$}
  & \colhead{$g$}
  & \colhead{t$_{buoy}$}
  & \colhead{t$_{r}$}
  & \colhead{t$_{c}$} \\ 
    \colhead{}
  & \colhead{}
  & \colhead{}
  & \colhead{(kpc)}
  & \colhead{(kpc)}
  & \colhead{(kpc)}
  & \colhead{(kpc)}
  & \colhead{(10$^{5}$ kpc$^{3}$)}
  & \colhead{(10$^{-8}$ cm s$^{-2}$)}
  & \colhead{(Myr)}
  & \colhead{(Myr)}
  & \colhead{(Myr)}
}
\startdata
RE      & 80$^{\text{o}}$ & N & 146 & 118 & 25 & 32 & 3.69 & 2.4 & 110 & 167 & 152 \\
RE      & 80$^{\text{o}}$ & S & 134 & 115 & 26 & 34 & 3.62 & 2.4 & 108 & 168 & 139 \\
RE      & 45$^{\text{o}}$ & N & 97  & 89  & 27 & 32 & 2.43 & 2.6 &  86 & 153 & 101 \\
RE      & 45$^{\text{o}}$ & S & 97  & 87  & 27 & 35 & 2.58 & 2.6 &  91 & 152 & 101 \\
RE      & 30$^{\text{o}}$ & N & 74  & 70  & 24 & 33 & 2.09 & 2.6 &  73 & 140 &  77 \\
RE      & 30$^{\text{o}}$ & S & 74  & 68  & 24 & 38 & 2.06 & 2.6 &  85 & 139 &  77 \\
I13     & 80$^{\text{o}}$ & N & 135 & 120 & 42 & 43 & 9.95 & 2.4 &  83 & 191 & 139 \\
I13     & 80$^{\text{o}}$ & S & 138 & 120 & 45 & 46 & 9.83 & 2.4 &  91 & 195 & 142 \\
I13     & 45$^{\text{o}}$ & N & 96  & 88  & 40 & 43 & 9.02 & 2.5 &  61 & 172 &  99 \\
I13     & 45$^{\text{o}}$ & S & 95  & 90  & 43 & 43 & 8.56 & 2.5 &  62 & 176 &  99 \\
I13     & 30$^{\text{o}}$ & N & 69  & 69  & 41 & 42 & 6.63 & 2.5 &  50 & 162 &  72 \\
I13     & 30$^{\text{o}}$ & S & 73  & 71  & 41 & 42 & 6.67 & 2.5 &  52 & 164 &  76 \\
\enddata
\tablecomments{Several cavity age estimates measured from observation for both RE and I13 at 157.5 Myr and 170.6 Myr respectively.}
\end{deluxetable}

\begin{deluxetable}{cccccc}
\tablecolumns{6}
\tablewidth{0pt}
\tablecaption{Cavity Power\label{tab:lum}}
\tablehead{
  \colhead{Model}
  & \colhead{$i$}
  & \colhead{H}
  & \colhead{$\langle t \rangle$}
  & \colhead{P$_{cav,obs}$}
  & \colhead{L$_{jet,act}$} \\
  \colhead{}
  & \colhead{}
  & \colhead{(10$^{60}$ erg)}
  & \colhead{(Myr)}
  & \colhead{(10$^{44}$ erg s$^{-1}$)}
  & \colhead{(10$^{44}$ erg s$^{-1}$)}
}
\startdata
RE      & 80$^{\text{o}}$ & 3.3  & 146 & 7.2   & 12 \\
RE      & 45$^{\text{o}}$ & 3.4  & 101 & 10.6  & 12 \\
RE      & 30$^{\text{o}}$ & 3.6  & 77  & 14.9  & 12 \\
I13     & 80$^{\text{o}}$ & 11.6 & 141 & 26.1  & 21 \\
I13     & 45$^{\text{o}}$ & 14.4 & 99  & 46.1  & 21 \\
I13     & 30$^{\text{o}}$ & 13.9 & 74  & 59.5  & 21 \\
\enddata
\tablecomments{Cavity power $P_{cav,obs}$ measured from observation for both 
RE and I13 at 157.5 Myr and 170.6 Myr respectively.  The average 
jet luminosity in the simulation out to those respective ages, $L_{jet,act}$, is given for comparison.}
\end{deluxetable}

\end{document}